\documentclass[a4paper]{article}
\usepackage{amsmath, amsfonts,  epsf, latexsym, cite, dsfont}

\numberwithin{equation}{section}
\def\balpha{\overline{\alpha}}
\def\bbeta{\overline{\beta}}

 \def\dem{\partial_{-}}
 \def\dep{\partial_{+}}
  \def\demm{\partial_{--}}
 \def\depp{\partial_{++}}

\newcommand{\pip}[1]{\pi_{+ #1}}
\newcommand{\pim}[1]{\pi_{- #1}}

\def\ext{\mathrm{ext}}
\def\min{\mathrm{min}}
\def\gh{\mathrm{gh}}

 \def\balpha{{\overline{\alpha}}}
 \def\bbeta{{\overline{\beta}}}
 \def\bgamma{{\overline{\gamma}}}

 \def\bkappa{{\overline{\kappa}}}

 \def\bnu{{\overline{\nu}}}

 \def\btau{{\overline{\tau}}}

\def\cM{{\cal{M}}}
\def\cD{{\cal{D}}}

\def\cO{{\cal{O}}}
\def\cP{{\cal{P}}}
\def\cJ{{\cal{J}}}
\def\cW{{\cal{W}}}

\def\hatSigma{\hat{\Sigma}}

\newcommand{\delr}{\raise.3ex\hbox{$\stackrel{\leftarrow}{\partial }$}}
\newcommand{\dell}{\raise.3ex\hbox{$\stackrel{\rightarrow}{\partial}$}}

\newcommand{\dr}{\raise.3ex\hbox{$\stackrel{\leftarrow}{\delta }$}}
\newcommand{\dl}{\raise.3ex\hbox{$\stackrel{\rightarrow}{\delta}$}}

{\catcode`\|=\active 
  \gdef\Braket#1{\left<\mathcode `\|"8000\let|\bravert {#1}\right>}}
\def\bravert{\egroup\,\vrule\, \bgroup}

\def\Jcpx{{\cal J}_{\mathrm{cpx}}}
\def\Jsym{{\cal J}_{\mathrm{sym}}}
 \def\Jpm{{\cal J}_{(\pm)} }
\def\Jp{{\cal J}_{(+)} }
\def\Jm{{\cal J}_{(-)}}

\def\ext{\mathrm{ext}}

\def\BRST{\mathrm{BRST}}
\def\BV{\mathrm{BV}}

\begin{document}
\thispagestyle{empty}
\begin{flushright}
\end{flushright}\vskip 0.8cm
\begin{center}
\LARGE{\bf Topological A-Type Models with Flux}
\end{center}
\vskip 1in

\begin{center}{\large Vid Stojevic }
\vskip 0.6cm{ II. Institut f\"{u}r Theoretische Physik der Universit\"{a}t Hamburg,\\
 Luruper Chaussee 149, 22761 Hamburg, Germany \\
 {\tt vid.stojevic@desy.de} }\\

\end{center}
\vskip 1.0cm

\begin{abstract}

\noindent

We study deformations of the A-model in the presence of fluxes, by which we mean rank-three tensors with antisymmetrized upper/lower indices, using the AKSZ construction.  There are two natural deformations of the A-model in the AKSZ language: 1) the Zucchini model, which can be defined on a generalized complex manifold and reduces to the A-model when the generalized complex structure comes from a symplectic structure, and 2) a topological membrane model, which naturally accommodates fluxes, and reduces to the Zucchini model on the boundary of the membrane when the fluxes are turned off.  We show that the fluxes are related to deformations of the Courant bracket which generalize the twist by a closed 3-from $H$, in the sense that  satisfying the AKSZ master equation implies precisely the integrability conditions for an almost generalized complex structure with respect to the deformed Courant bracket.  In addition, the master equation imposes conditions on the fluxes that generalize $dH=0$. The membrane model can be defined on a large class of $U(m)$- and $U(m) \times U(m)$-structure manifolds relevant for string theory, including geometries inspired by $(1,1)$ supersymmetric $\sigma$-models with additional supersymmetries due to almost complex (but not necessarily complex) structures in the target space.  In addition we show that the model can be defined on three particular half-flat manifolds related to the Iwasawa manifold.

When only the closed 3-form flux is turned on it is possible to obtain a  topological string model, which we do for the  case of a Calabi-Yau. We argue that deformations from the standard A-model are due to the choice of gauge fixing fermion, rather than a flux deformation of the AKSZ action. The particularly interesting cases arise when the fermion depends on  auxiliary fields, the simplest possibility being due to the $(2,0)+ (0,2)$ component of a non-trivial $b$-field.  The model is generically no longer evaluated on holomorphic maps and defines new topological invariants. Deformations due to $H$-flux can be more radical,  completely preventing auxiliary fields from being integrated out.

\end{abstract}

\vfill

\setcounter{footnote}{0}
\def\thefootnote{\arabic{footnote}}
\newpage

\section{Introduction}

In this paper we study deformations of the topological A-model  \cite{Witten:1988xj} by rank-three tensors with anti-symmetrized upper/lower indices, $H_{ijk}$, $f^i_{ \ jk}$, $Q^{ \ jk}_i$, and $R^{ijk}$, which we refer to as fluxes.  Our approach makes use of the AKSZ construction, originally introduced in \cite{Alexandrov:1995kv} by  Alexandrov, Kontsevich, Schwarz, and Zoboronsky, which is a very general framework that enables a geometric construction of topological models by making use of the Batalin-Vilkovisky (BV) formalism   \cite{Batalin:1981jr, Batalin:1984jr}. Because the construction involves standard gauge fixing techniques, it is conceptually very different to the 'twisting' procedure, where one obtains a topological theory starting from a $(2,2)$ supersymmetric $\sigma$-model, but the BRST operator has no interpretation as originating from some underlying gauge theory. In the AKSZ construction deformations from the A-model are naturally introduced in two stages. First we consider the standard A-model action,  and deform the symplectic structure to some generalized complex structure \cite{Gualtieri:2003dx}. In this step one is essentially concerned with the Zucchini model \cite{Zucchini:2004ta} on a generalized complex manifold.\footnote{The Zucchini model can also be defined on more general "Poisson-quasi-Nijenhuis" geometries \cite{Zucchini:2007ie}, but we do not consider such geometries in this paper.} In the second stage we consider an AKSZ action of a topological membrane with a boundary, which naturally incorporates the fluxes and reduces to the Zucchini model on the boundary when these are turned off. When the fluxes are turned on the membrane Lagrangian is no longer a total derivative, so the theory is genuinely three dimensional.   Topological membrane models correspond to Courant algebroids \cite{LWX}, and have been studied extensively in the mathematics  \cite{Roy1, Roytenberg:2002nu, Roytenberg:2006qz, Kotov:2004wz, Alekseev:2004np} and the physics literature \cite{Ikeda:2002wh, Ikeda:2002qx, Ikeda:2006wd, Ikeda:2006pd, Ikeda:2007rn}.   

In order to define a topological model the BV master equation must be satisfied, which implies constraints on the target space geometry. Without fluxes the master equation is satisfied for a generalized complex geometry. When the fluxes are turned on we show that the generalization can be understood in terms of the integrability conditions for an almost generalized complex structure with respect to a deformation of the standard Courant bracket by the $H$-, $f$-, $Q$- and $R$-fluxes (when only the $H$-flux is turned on this is the standard $H$-twisted bracket studied  in \cite{Gualtieri:2003dx}).  More precisely, there are two groups of equations that arise from the master equation. The first group is satisfied if the integrability conditions with respect to the deformed bracket are, and the second group contains Bianchi identity type relations which generalize the condition $dH=0$. The latter condition arises by requiring the $H$-twisted bracket to be a Courant algebroid bracket, but the differential conditions on the $f$-, $Q$-, and $R$-fluxes can not be understood in an analogous manner. 

The BV master equation can be satisfied for a large class of almost generalized complex manifolds with flux, some of which are relevant for string theory. We study the topological membrane on almost complex $U(m)$-structure geometries related to $(2,1)$-supersymmetric $\sigma$-models, for which the torsion is obtained by raising an index of a closed 3-form $H$.  One can also consider $U(m)$ geometries for which the torsion is not totally antisymmetric.  Either way, it is the $f$-flux, given by the torsion tensor, that needs to be turned on in the topological model, and the Bianchi identity type condition can be written as $R^{(T)}_{ \btau [ \alpha \beta \gamma]} = 0$, where $R^{(T)}$ is the torsionful Riemann tensor. This is automatically satisfied for the $\sigma$-model geometries, but provides a constraint when the torsion is not totally antisymmetric, which we show is satisfied for three explicit examples of half-flat manifolds.  Furthermore, we discuss a class of geometries related to $(2,2)$ supersymmetric $\sigma$-models. These have $U(m) \times U(m)$ structure, and are characterized by two almost complex structures, one of which is integrable and the other not. The fluxes that need to be turned on in the topological model are $H$, $f$, and $R$, where $f$ and $R$ are obtained by raising the indices of $H$. The $\sigma$-model geometries are expected to contain solutions of the string theory equations of motion to lowest order in $\alpha'$ \cite{Grana:2005jc}, in a warped compactification to flat spacetime, while half-flat manifolds arise as mirrors of Calabi-Yau three-folds with electric NS 3-form flux \cite{Gurrieri:2002wz}.  In relation to type II  compactifications to four dimensions, we argue that the $SU(3)$ $\sigma$-model geometries should break all spacetime supersymmetry, and the $SU(3) \times SU(3)$ geometries leave $N=1$ supersymmetry intact. 

To fully specify the topological theory one must impose a gauge fixing prescription, and this step is in general very intricate. We analyze the gauge fixing only in the context of two-dimensional models with $H$-flux, by exploiting the fact that in this case it is possible to obtain a string model from the membrane model. This is because one can always perform an appropriate $b$-transform in the membrane model, with $db\propto H$, in order to obtain an action that is a total derivative and reduces to a string model on the boundary.  We study a Calabi-Yau manifold with $3$-form flux in this setting. Gauge fixing follows directly from the standard A-model, and  we argue that deformations from the A-model actually occur at the level of gauge fixing. This seems to contradict the formal proofs in BV quantization which show an anomaly free theory to be invariant under deformations of the gauge fixing fermion. However, these proofs don't take into account non-perturbative effects, which are central for topological theories. Non-trivial deformations of the A-model can occur when the gauge fixing fermion depends on $H$-flux or the $(2,0) + (0,2)$ component of a non-trivial $b$-field. In general this means that the model is no longer evaluated on holomorphic maps, and describes new topological invariants.  The $b$-field case is the simplest, and we study it in detail. Deformations due to $H$-flux can have a more dramatic impact, preventing fields which are auxiliary in the AKSZ construction of the standard A-model from being integrated out.

The outline of the paper is as follows. In Sections \ref{sec:gen_cpx_geom} and \ref{sec:BV_procedure}  we give brief reviews of generalized complex geometry and the BV procedure. The deformed Courant bracket is introduced in Section  \ref{sec:gen_cpx_geom}, and the integrability conditions for an almost generalized complex structure with respect to this bracket are given in Appendix  \ref{app:integrability}. The AKSZ construction is reviewed in Section \ref{sec:AKSZ_construction}, with particular emphasis on the A-model and the Zucchini model. In Subsection \ref{sec:inv_under_b_transforms} we demonstrate the invariance of the A-model under $b$-transforms (with $db=0$, unlike in the context of a Calabi-Yau with flux above), showing that the effect of a $b$-transform can be undone by an appropriate choice of a gauge fixing fermion, but only once the auxiliary fields have been integrated out. In Section \ref{sec:almost_cpx_models} we study topological membrane models on $U(m)$- and $U(m) \times U(m)$-structure manifolds with almost complex structures, and in Section \ref{sec:CY_with_flux} we study topological string models on Calabi-Yau manifolds with 3-form flux. Some concluding remarks are given in Section \ref{sec:conclusions}.

\section{Generalized Complex Geometry}
\label{sec:gen_cpx_geom}

An almost generalized complex structure \cite{Gualtieri:2003dx}  on a manifold $\cM$ is an an endomorphism $ {\cal J}$ of the sum of the tangent and cotangent bundles $ T \cM \oplus T^* \cM$ which satisfies  ${\cal J}^2 = - \mathds{1}$, and is  compatible with the canonical metric  ${\cal G}$ on $T \cM \oplus T^* \cM$:
\begin{equation}
\label{eq:gen_cpx_metric}
 {\cal G} = \left ( \begin{array}{ll}
                      0  & \mathds{1} \\
                      \mathds{1}  & 0
                    \end{array} \right ) \ .
\end{equation}
This generalizes the concept of an almost complex structure $I$, which is an endomorphism of $T \cM$ that squares to $-1$.  
Writing $ {\cal J} $ in components\footnote{We roughly follow the component notation of \cite{Lindstrom:2004iw}.},
\begin{equation}
\label{eq:gen_cpx_structure}
 {\cal J} = \left ( \begin{array}{ll}
                      J  & P \\
                      L  & K
                    \end{array} \right ) \ ,
\end{equation}
 the compatibility with respect to ${\cal G}$ reads
\begin{align}
\label{eq:gen_cpx1}
J^{i}_{ \ j} + K^{ \ j}_{i} = 0 \ \ \ ,  \ \  \  L_{ij} + L_{ji} = 0 \ \ \ \mathrm{and} \ \ \ P^{ij} + P^{ji} = 0 \ ,
\end{align}
and ${\cal J}^2 = - \mathds{1}$ is equivalent to
\begin{align}
\label{eq:gen_cpx2}
J^k_{ \ i} J^i_{ \ p} + P^{ki} L_{ip}=  - \delta^k_p  \ \ \ , \ \ \  J^{(k}_{ \ i} P^{j)i} = J^i_{ ( j} L_{k) i} = 0 \ .
\end{align}

Using an almost complex structure one can define the operators
\begin{equation}
p_{\pm} := \frac{1}{2} ( \mathds{1} \pm i I) 
\end{equation}
that project onto holomorphic and antiholomorphic subspaces of $T \cM \otimes \mathbb{C}$. If these are involutive with respect to the Lie bracket the almost complex structure is integrable. The natural generalization of the Lie bracket for $T \cM \oplus T^* \cM$ is the Courant bracket,
\begin{align}
\label{eq:Courant_bracket}
[X+ \eta, Y+ \nu]_C  =  [X, Y] + \mathcal{L}_X \nu  - \mathcal{L}_Y \eta - \frac{1}{2} d ( i_X \nu - i_Y \eta ) \ ,
\end{align}
where $X, Y $ are sections of $T \cM$ and $\eta,  \nu$ sections of $T^* \cM$. An almost generalized complex structure is actually generalized complex if the eigenbundles $l$ and $l^*$ of  $(T \oplus T^*) \otimes \mathbb{C}$ defined by the projectors
\begin{equation}
\Pi_{\pm} : = \frac{1}{2} ( \mathds{1}  \pm i \cal J)  \ 
\end{equation}
are involutive  with respect to the Courant bracket:
\begin{equation}
\label{eq:int_conditions}
\Pi_{\mp} [ \Pi_{\pm} (X+ \eta), \Pi_{\pm} (Y+ \nu) ]_{C} = 0 \ .
\end{equation}
One can show that $l$ and $l^*$  are maximal isotropic subbundles of $(T \cM \oplus T^* \cM) \otimes \mathbb{C}$, and are indeed dual spaces, as the notation indicates. The integrability of an almost generalized complex structure with respect to the Courant bracket implies that one can work in a chart which is a product of the standard complex space and the standard symplectic space.\footnote{This result is derived in section 4.7 of \cite{Gualtieri:2003dx}.} 

Complex and symplectic geometries are special cases of the integrable generalized complex structures
\begin{align}
\label{eq:sym_and_cpx}
\Jcpx = \left ( \begin{array}{ll}
                      I  & 0  \\
                      0  & -I^t
                    \end{array} \right )  \ \ \  \mathrm{and} \ \ \  \Jsym = \left ( \begin{array}{ll}
                      0  & - \omega^{-1} \\
                      \omega  & 0
                    \end{array} \right ) \ ,
\end{align}
where $\omega$ is a symplectic form. K\"{a}hler manifolds hold both of these structures, in which case $\omega$ is the K\"{a}hler form. A less trivial example is that of hyper-K\"{a}hler geometry, characterized by a pair of  covariantly constant complex structures $I_{(\pm)}$, from which the two generalized complex structures
\begin{equation}
\label{eq:bi_hermitian_gen_cpx0}
\Jpm = \left ( \begin{array}{ll}
                      I^{ \ \ i}_{(+) j} \pm I^{ \ \ i}_{(-) j}  &  -(I_{(+)}^{ij} \mp I_{(-)}^{ij})\\
                      I_{ij}^{(+)} \mp I_{ij}^{(-)}    & -  ( I^{ (+) i}_{  \ \ j} \pm  I^{ (-) i}_{ \ \  j})
                    \end{array} \right ) \ 
\end{equation}
can be constructed.

The structure group of $T \cM \oplus T^* \cM$ that preserves $ {\cal G}$ is $O(d,d)$, where $d$ is the dimension of $\cM$. The subgroup of $O(d,d)$ that leaves the Courant bracket invariant is given by diffeomorphisms and $b$-transform, the latter being skew $O(d,d)$ transformation that act as
\begin{equation}
\label{eq:b_transform}
e^b(X+ \eta) = X + \eta + i_X b \ ,
\end{equation}
for some closed 2-form $b$. Under a $b$-transform by a $b$ field that is not closed, but obeys $3db = H$, the Courant bracket is deformed to the H-twisted Courant bracket:
\begin{equation}
\label{eq:H_twisted_courant}
[X+ \eta, Y+ \nu]_H =  [X+ \eta, Y+ \nu]_C + i_X i_Y H \ .
\end{equation}
It turns out that the interpretation of the integrability conditions with respect to $[ , ]_H$ are the same as for the standard Courant bracket, the crucial property in the derivation being that $b$-transforms don't change the type of the generalized complex structure - type being the dimension of the complex subspace. Bi-Hermitian geometry \cite{Gates:1984nk, Gualtieri:2003dx, Lindstrom:2004hi, Lindstrom:2005zr, Bredthauer:2006hf, Lindstrom:2006ee, Lindstrom:2007qf, Lindstrom:2007xv, Lindstrom:2007sq} is characterized by $\Jpm$ integrable with respect to $[,]_H$, while a generalization of K\"{a}hler geometry with torsion \cite{Hull:1985jv, Hull:1997kk} is characterized by $\Jcpx$ and $\Jsym$ integrable with respect to $[,]_H$.  We will give more details about both of these geometries in Section \ref{sec:almost_cpx_models}.

The $+i$ eigenbundle $l$ has the structure of a \emph{Lie algebroid} \cite{Gualtieri:2003dx, Mackenzie}. A Lie algebroid $L$  is a vector bundle on $\cM$ together with a Lie bracket $[ \cdot , \cdot ]$ that acts on sections of $L$, and an anchor map $a: C^{\infty} (L) \rightarrow C^{\infty} (T \cM)$ that satisfies:
\begin{align}
\label{eq:Lie_algebroid}
& a ( [ X,Y]) = [a(X) , a(Y) ] \\ \nonumber
& [X, fY ] = f [ X, Y] + (a (X) f) Y \ \ \ \forall X, Y \in C^{\infty} (L), f \in C^{\infty} (\cM) \ .
\end{align} 
There is a natural exterior derivative $d_L: C^\infty(\Lambda^k L^*) \rightarrow C^\infty(\Lambda^{k+1} L^*)$, where $L^*$ is the dual space to $L$,  that can be defined using the Lie bracket and obeys $(d_L)^2=0$. Furthermore, $l \oplus l^*$ has the structure of a \emph{Lie bi-algebroid}, which is a pair of Lie algebroids $(L, L^*)$ with the  additional requirement that $d_L$ obeys the Leibnitz rule  with respect to the Lie algebroid bracket on $L^*$. 

The underlying reason why $l \oplus l^*$ is a Lie bi-algebroid is that the ($H$-twisted) Courant bracket belongs to a \emph{Courant algebroid} structure.  A Courant algebroid \cite{LWX, Roy1, Gualtieri:2003dx}  is a vector bundle $E$ with a bilinear form $\langle , \rangle$, a bracket $[ , ]$, and an anchor map $\pi: E \rightarrow T \cM$ obeying
\begin{align}
\label{eq:def_Courant_alg}
& \pi ( [ X,Y]) = [ \pi (X) , \pi (Y) ]  \ \ \ \forall X, Y \in C^{\infty} (E), \\ \nonumber
& [X, fY ] = f [ X, Y] + (a (X) f) Y + (\pi(X) f) Y - \langle X, Y \rangle \cD f \ \ \ \forall X, Y \in C^{\infty} (E), f \in C^{\infty} (\cM) \\ \nonumber
& \langle \cD f, \cD g \rangle = 0 \ \ \ \forall f, g \in C^{\infty} (\cM)  \\ \nonumber 
& \pi (X) \langle Y, Z \rangle - \langle [X, Y ] + \cD \langle X, Y \rangle, Z \rangle + \langle Y, [X, Z] + \cD \langle X, Z \rangle \rangle \ \ \ \forall X, Y, Z \in C^\infty (E)  \\ \nonumber 
& [X, [Y, Z]] + [Y, [Z, X]] + [Z, [X,Y]] = \frac{1}{3} \cD (  \mathrm{Nij} (X, Y, Z) ) \ \ \  \forall X, Y, Z  \in C^{\infty} (E)  \ ,
\end{align} 
where
\begin{equation}
\mathrm{Nij} (X, Y, Z) :  =  \langle [X, Y], Z \rangle +  \langle [Y, Z], X \rangle  +  \langle [Z, X], Y \rangle \ ,
\end{equation}
and $\cD$ is a map $C^\infty(\cM) \rightarrow C^\infty(E) $ defined by the property $\langle \cD f, X \rangle = \frac{1}{2} \pi (X) f$  \ $\forall f \in C^\infty(\cM),  X \in C^\infty (E)$.  The Lie brackets on $l$ and $l^*$ are obtained by restricting the ($H$-twisted) Courant bracket, and the anchor maps are given by restricting the projection $(T \cM \oplus T^* \cM) \rightarrow T \cM$. The important property is the last one in (\ref{eq:def_Courant_alg}), because the restriction of an  ($H$-twisted)  Courant algebroid bracket to an involutive maximal isotropic subspace obeys the Jacobi identity, and is therefore a Lie bracket. 

In this paper we will construct topological membrane models on almost generalized complex geometries that obey integrability conditions with respect to brackets obtained by general deformations of the Courant bracket by rank three tensors\footnote{Twists by fluxes other than $H$ have been considered in the formulation of generalized complex geometry in terms of $O(n,n)$ pure spinors in \cite{ Grana:2004bg, Grana:2005sn, Grana:2006hr, Grange:2007bp}. }:
\begin{align}
\label{eq:courant_def}
[X+ \eta, Y+ \nu]_{D} =  &  [X+ \eta, Y+ \nu]_C + k_1 H_{k ij} X^i Y^j + k_2 f^k_{ \ ij} X^i Y^j \\ \nonumber
& - k_2 f_{ \ ik}^{ j} (X^i \nu_j - Y^i \eta_j) - k_3 Q^{ \  j k }_{ i} (X^i \nu_j - Y^i \eta_j)  \\ \nonumber
& + k_3 Q_k^{ \ ij} \eta_i \nu_j + k_4 R^{kij} \eta_i \nu_j \  ,
\end{align}
where $\{ k_1, \cdots, k_4 \}$ is a set of parameters, and  the position of the free indices denotes whether the term lives in $T \cM$ or $T^* \cM$.  The explicit integrability conditions for a general $\cal{J}$ with respect to this bracket are given in Appendix \ref{app:integrability}.  The bracket $[ , ]_D$ is not a Courant algebroid bracket if any combination of $f$, $Q$, and $R$ is turned on, which is related to the fact that these fluxes can not be understood as coming from some kind of a transform analogous to (\ref{eq:b_transform}). For example, whereas
\begin{equation}
\label{eq:b_transf_of_courant}
[e^b(X+ \eta), e^b(Y+ \nu)]_C  = e^b [X+ \eta, Y+ \nu]_C  + i_Y i_X db \equiv  e^b [X+ \eta, Y+ \nu]_H \ ,
\end{equation}
for a transformation by a bivector $\beta$,
\begin{equation}
e^\beta( X + \eta) = X + \eta + i_\eta \beta \ ,
\end{equation}
one doesn't have an analogue of (\ref{eq:b_transf_of_courant}). That is,
\begin{equation}
[e^\beta (X+ \eta), e^\beta (Y+ \nu) ]_C  \neq e^\beta [X+ \eta, Y+ \nu]_D \ 
\end{equation}
irrespective of which fluxes in $[ \cdot, \cdot ]_D$ are turned on. One can also check that unless at most the $H$ flux is turned on, $[ \cdot, \cdot ]_D$ is not a Courant algebroid bracket, in which case the maximal isotropic subspaces of $(T \cM \oplus T^* \cM) \otimes \mathbb{C}$ involutive with respect to $[ , ]_D$ are not Lie algebroids, and the generalized Darboux theorem is no longer valid. Never the less, as we discuss in Section  \ref{sec:3dAtype_model}, a topological model on a target spaces with an almost generalized complex structure integrable with respect to $[ , ]_D$ corresponds to a Courant algebroid structure on $\cM$. The topological model also places differential constraints on the fluxes that resemble Bianchi identities. In the case of $H$-flux this is simply $dH=0$, a condition that is independent of any almost generalized complex structure and follows from the requirement that  $[,]_H$ be a Courant algebroid bracket. The conditions on the other fluxes do not seem to follow from a property of $[ , ]_D$ alone. We emphasize that the integrability conditions with respect to $[ , ]_D$ are of importance to this paper because they appear as consistency conditions for topological models with flux, but we do not claim to understand the full mathematical significance of $[ , ]_D$ at this stage.

\section{ Batalin-Vilkovisky formalism}
\label{sec:BV_procedure}

The Batalin-Vilkovisky (BV) formalism is a general recipe for gauge fixing in the Lagrangian framework, which on one hand provides a procedure to obtain a gauge-fixed action starting from a classical gauge theory, and furthermore gives a condition, known as the quantum master equation, that must be satisfied in order for the observables and the partition function to be independent of the gauge choice.  It was introduced in \cite{Batalin:1981jr, Batalin:1984jr}, and enabled the quantization of gauge theories with open and reducible algebras, which could not be handled using the BRST techniques available at the time. BV quantization encompasses BRST, and improves over it by uncovering the geometric structures involved in the gauge fixing procedure.  As we will show in detail, the construction of a gauge fixed action involves doubling the field space by introducing a partner with opposite statistics, referred to as an antifield, for every physical field and ghost field in the theory. The field-antifield space has a canonical odd symplectic structure on it. In addition, an extended action $S$ that is nilpotent with respect to the odd symplectic structure needs to be constructed. A supermanifold with an odd symplectic structure and a nilpotent vector field defines a BV geometry, and it is in fact not necessary to start the construction from a classical gauge invariant action, since any BV geometry will serve as a starting point for the definition of a quantum theory. In particular, in the AKSZ procedure BV geometries are constructed directly from classical geometries. In this section we review the standard approach to BV quantization\footnote{For a thorough treatment the reader is referred to the review \cite{Gomis:1994he}, or to \cite{DeJonghe:1993zc, Vandoren:1996ku}.}, and in the next we will address the AKSZ approach.  This will provide a physical motivation for BV geometry, and is also necessary for understanding the content of the topological theories constructed via AKSZ. 

Suppose we have an action $S_0$, that depends on a set of fields $\phi^i$ and is invariant under a set of gauge symmetries labeled by a capital letter index,\footnote{Throughout this section we will be using the deWitt convention, where repeated indices imply not only summation but also integration. In (\ref{eq:gauge_transf}), for example, the transformation parameters $\varepsilon^A$ are functions, while $R_A^i$ are really object that contain a ($d$-dimensional) $\delta$-function, so:
\begin{equation}
\varepsilon^A R_A^i  \equiv \int dz \varepsilon^A(z)  R_A^i(x-z) \ .
\end{equation}
In general, an object with $n$ free indices in the deWitt notations stands for something with $(n-1)$ $\delta$-functions; for $n=0$ we have a local integrated object. The deWitt notation is very useful when discussing the BV formalism abstractly, without reference to a particular theory, but we will not use when discussing specific constructions later in the paper. }
\begin{equation}
\label{eq:gauge_transf}
\delta \phi^i = \varepsilon^A R_A^i \ ,
\end{equation}
so,
\begin{equation}
\label{eq:inv_of_action}
\frac{\dr S_0}{ \delta \phi^i} \varepsilon^A R_A^i =0 \ .
\end{equation}
Gauge symmetries imply that the Hessian,
\begin{equation}
H_{ij} := \frac{\dl}{\delta \phi^i} \frac{\dr S_0}{ \delta \phi^j}
\end{equation}
is not invertible on-shell, which can be seen by taking a functional derivative of (\ref{eq:inv_of_action}).  This prevents the construction of the free propagator (given by the inverse of the Hessian evaluated at some classical solution), and thus prevents a perturbative evaluation of the path integral. 

As in BRST, one introduces ghost fields $c^A$, characterized by having opposite parity to the gauge transformation parameters $\varepsilon^A$. We group the ghosts together with the original $\phi^i$ under a collective field,
\begin{equation}
\label{eq:collective_field}
\Phi^\alpha = \{ \phi^i, c^A \} \ ,
\end{equation}
and for each $\Phi^\alpha$ introduce a field with opposite statistics $\Phi^*_\alpha$, called an antifield. Then an extended action starting as
\begin{equation}
\label{eq:min_solution}
S_\min = S_0 + \phi^*_i c^A R^i_A + \cdots 
\end{equation}
 is constructed. The dots are completed by requiring $S_\min$ to be a solution to the master equation
\begin{equation}
\label{eq:master_equation}
(S_\min, S_\min) = 0 \ ,
\end{equation}
where $( \cdot, \cdot)$ is called the antibracket, and defines a canonical odd symplectic structure in the space of fields and antifields (see Appendix \ref{app:poisson_antibracket} for a list of defining properties of the antibracket). Its action on two objects $A$ and $B$ that depend on $\Phi$ and $\Phi^*$ is given by
\begin{equation}
\label{eq:antibracket_def}
(A, B) :=   \frac{ \dr A} { \delta   \Phi^\alpha } \frac{\dl B}{ \delta \Phi^*_\alpha} - 
\frac{\dr A}{ \delta \Phi^*_\alpha }  \frac{ \dl B}{ \delta   \Phi^\alpha }  \  .
\end{equation}
For reasons that will become clear shortly, $S_\min$ is called the minimal solution, and is the central ingredient for constructing a gauge-fixed action.

For a symmetry algebra that is not reducible and closes on-shell the solution to the master equation is given by
\begin{equation}
S_{\min} = S_0 + \phi^*_i c^A R_A^i + c^*_C N^C_{ \ AB} c^A c^B \ ,
\end{equation}
where $N^C_{ \ AB}$  are the (possibly field dependent) structure functions of the algebra. The master equation reads
\begin{align}
\frac{1}{2} (S_\min ,S_\min ) =  & \frac{ \dr S_0 }{\delta \phi^i} c^A R^i_A   + \phi^*_i \left[ c^A \frac{ \dr  R^i_A}{\delta \phi^k} c^B R^k_B + (-1)^{\epsilon_{\phi_i} \epsilon_{c^F}} R^i_F N^F_{ \ AB} c^A c^B \right] \\ \nonumber
& + c^*_D \left[2 N^D_{ \ AF} c^A N^F_{ \ GH} c^G c^H  - c^B c^C \frac{\dr N^D_{BC} }{\delta \phi^k} c^A R^k_A \right] = 0 \ ,
\end{align}
where $\epsilon \in \mathbb{Z}_2$ is zero when the field in its subscript is bosonic and one when it is fermionic. The term independent of antifields expresses the invariance of the action, the term proportional to $\phi^*$ the closure of the algebra, while the term proportional to $c^*$ is related to the Jacobi identity\footnote{There is a technical issue here, that rarely causes an obstruction to the BV procedure in practice \cite{Howe:2006si, Stojevic:2006pq}. The Jacobi identity associated to a set of gauge transformations (\ref{eq:gauge_transf}) that close on-shell is 
\begin{equation}
R^i_D \left( N^D_{ \ AF} c^A N^F_{ \ GH} c^G c^H - c^B c^C \frac{\dr N^D_{BC} }{\delta \phi^k} c^A R^k_A \right)  \equiv 0  \ .
\end{equation} Therefore, this relation must be satisfied already at the level of the structure functions for the BV procedure to work.}.   Essentially, the antifields act as sources for the BRST transformations obtained when quantizing a gauge theory with a closed and irreducible gauge algebra using the standard BRST methods,
\begin{align}
& \delta_\BRST \phi^i : = \theta c^A R_A^i  \ \ \ , \ \ \ \delta_\BRST c^C : = \theta N^C_{ \ AB} c^A c^B  \ ,
\end{align}
where $\theta$ is a global fermionic transformation parameter. The statement that $\delta_\BRST$ is nilpotent is equivalent to the statement that (\ref{eq:me_with_aux}) is a solution to the master equation. For open algebras, when the BRST procedure fails, the solution to the master equation is characterized by terms non-linear in the antifields. Still, it is customary to refer to the transformations that appear in the terms linear in antifields as BRST transformations, even though for open algebras these transformations are nilpotent only on-shell. 

Gauge fixing is achieved via a canonical transformation, that is, a transformation that preserves the antibracket (\ref{eq:antibracket_def}). An infinitesimal canonical transformation can be generated by a fermionic function $f(\Phi, \Phi^*)$ of fields and antifields as
\begin{equation}
\Phi^*_\alpha \rightarrow \Phi^*_\alpha + (f(\Phi, \Phi^*), \Phi^*_\alpha ) \ \ \ ,  \ \ \ \Phi^\alpha \rightarrow \Phi^\alpha + (f(\Phi, \Phi^*) , \Phi^\alpha) \ ,
\end{equation}
and it turns out that for the purposes of gauge fixing we can restrict to functions depending only on fields, i.e. to canonical transformations acting only on antifields:
\begin{equation}
\label{eq:gauge_fixing}
\Phi^*_\alpha \rightarrow \Phi^*_\alpha + (\Psi(\Phi), \Phi^*_\alpha ) \ \ \ ,  \ \ \ \Phi^\alpha \rightarrow \Phi^\alpha  \ .
\end{equation}
$\Psi$ is referred to as the \emph{gauge fixing fermion}. The minimal solution has a ghost number symmetry, where the ghost numbers are conventionally assigned as
\begin{equation}
\gh (\phi^i) = 0 \ \ \ , \ \ \ \gh (c^A) = 1 \ \ \ ,  \ \ \ \gh (\Phi^*_\alpha) = - \gh (\Phi^\alpha) - 1 \ .
\end{equation}
The antibracket increases ghost number by one, so in order for the canonical transformation to preserve ghost number it's necessary that $\gh (\Psi) = -1$. Such a fermion can't be constructed from the fields in the minimal solution, since these all have positive ghost number. One therefore introduces auxiliary field pairs $b^A$ and $\lambda^A$ with $\gh (b^A) = -1$ and $\gh (\lambda^A) = 0$, which can be used to construct an appropriate $\Psi$. The extended action with the auxiliary fields reads
\begin{equation}
\label{eq:me_with_aux}
S_{\ext} = S_\min + b^*_A \lambda^A \ ,
\end{equation}
so on the auxiliaries the BRST transformation simply acts as 
\begin{align}
& \delta_\BRST b^A = \lambda^A  \ \ \ , \ \ \ \delta_\BRST \lambda^A = 0 \ .
\end{align}
For closed, irreducible algebras, gauge fixing via a canonical transformation (\ref{eq:gauge_fixing}) is equivalent to adding the term $\delta_\BRST \Psi$ to $S_\ext$, but for open algebras terms nonlinear in antifields play a crucial role. 

For irreducible algebras it is always possible to find a fermion such that the part of the transformed extended action independent of antifields has a propagator. As things stand, the procedure still fails for reducible algebras. This is fixed by requiring the minimal solution to be \emph{proper}, which means that the Hessian defined by
\begin{equation}
H_{ab} := \frac{\dl}{\delta \varphi^a} \frac{\dr S_{\ext} }{\delta \varphi^b}
\end{equation}
has half its maximal rank on-shell. Here $\varphi^a$ is a collective field that includes $\phi^i$, $c^A$, any other ghost fields that need to be included for the solution to the master equation to be proper, as well as the antifields for all these. One can show that for irreducible algebras a minimal solution with the field content (\ref{eq:collective_field}) that starts  as (\ref{eq:min_solution}) is indeed proper,  while for reducible algebras additional ghost fields need to be introduced.  The properness condition guarantees that, after a canonical transformation,  one can find an action with a well defined free propagator.

To summarize, the object that can be used as the starting point for defining the quantum theory is
\begin{equation}
S_\ext (\Phi^\alpha, \Phi^*_\alpha + (\Psi(\Phi), \Phi^*_\alpha ) ) \ ,
\end{equation}
where $\Phi^\alpha$ now includes the auxiliaries $b^A$ and $\lambda^A$. Furthermore, one can show that the generating functional
\begin{equation}
Z[J, \Phi^*_\alpha] = \int [d \Phi] \exp{i (W_\ext + \Phi^\alpha J_\alpha)}
\end{equation}
obeys the \emph{naive Ward identity}
\begin{equation}
\label{eq:ward_identity1}
J_\alpha \frac{\dl Z[J, \Phi^*_\alpha] }{\delta \Phi^*_\alpha} = 0 \ ,
\end{equation}
provided that $W_\ext$ obeys the quantum master equation
\begin{equation}
\label{eq:quantum_me2}
-i \Delta W_\ext + \frac{1}{2} (W_\ext , W_\ext ) = 0 \ ,
\end{equation}
where the $\Delta$ operator is defined as
\begin{equation}
\label{eq:Delta_operator}
\Delta := (-1)^{(\epsilon_i + 1)} \frac{\dr}{ \delta \Phi^*_i } \frac{ \dr}{\delta \Phi^i} \ .
\end{equation}
By 'naive' we mean that path integral quantities are manipulated as if they were classical objects, without regard for regularization and renormalization; the 'naive' results can be spoiled by quantum anomalies. In particular, $W_\ext$ should be thought of as depending on some regularization parameter $\epsilon$, such that when the limit $\epsilon \rightarrow 0$ is taken it becomes infinite and $Z[J]$ finite. On the other hand the $\Delta$ operator is singular when acing on local functionals, and must be regularized if one is to make sense of equation (\ref{eq:quantum_me2}). In the above we have set $\hbar = 1$; if  $\hbar$ was reintroduced the $\Delta$ operator  term in (\ref{eq:quantum_me2}) would come out proportional to $\hbar$, and should therefore be understood as a quantum correction.\footnote{The $\Delta$ operator is avoided when working with the effective action,
\begin{equation}
\label{eq:effective_action_me2}
\Gamma [ \Phi^i_{(c)}, \Phi^*_\alpha] := -i Z[J, \Phi^*]  - J_\alpha \Phi^\alpha_{(c)} \ .
\end{equation}
where
\begin{equation}
 \Phi^\alpha_{(c)}  := \frac{ -i \dr \ln Z}{\delta J_\alpha}
\end{equation}
Then (\ref{eq:ward_identity1}) is equivalent to the effective action obeying the classical master equation
$( \Gamma, \Gamma)_{\Phi_{(c)}} = 0$, where the subscript indicates that the functional derivatives are with respect to $\Phi^i_{(c)}$.}

To demonstrate that the BV gauge fixing procedure makes sense one needs to show that the quantum theory is invariant under infinitesimal deformations of the gauge fixing fermion. And indeed one can argue, naively, that the partition function is invariant provided that (\ref{eq:quantum_me2}) is satisfied (see any of \cite{Gomis:1994he, DeJonghe:1993zc, Vandoren:1996ku} for the details). Furthermore, one can show that the expectation value of an observable $\cO$ is independent of the choice of $\Psi$ provided that 
\begin{equation}
\label{eq:observables}
\sigma(\cO)   = 0 \ ,
\end{equation}
where $\sigma$ is defined by
\begin{equation}
\sigma \cO : = (\cO, W_\ext) - i \Delta \cO \ ,
\end{equation}
and formally obeys $\sigma^2=0$ (clearly, this reduces to the invariance of the partition function itself since $\sigma W_\ext = 0$ is just the quantum master equation).  Due to the nilpotence of $\sigma$, (\ref{eq:observables}) is automatically true when $\cO$ is of the form $\cO = \sigma \mathcal{F} $.  It follows that quantum observables are given by the cohomology classes of  $\sigma$ at ghost number zero, where the ghost number requirement comes from considering the classical limit. In the classical limit the $\Delta$ operator term vanishes, so $\sigma$ reduces to the operator $\delta_{\BV}$,
\begin{equation}
\label{eq:BVoperator}
\delta_{\BV} A : = (A, W_\ext) \ ,
\end{equation}
whose nilpotence follows directly from the Jacobi identity property of the antibracket (\ref{eq:ab_jacobi}). The cohomology of $\delta_{\BV}$ restricted to local functionals determines the deformations of the classical action compatible with the symmetries of the theory (which in particular determines potential counterterms). Using the fact that $\gh(\delta_{\BV}) = 1$, one can also show that the cohomology of local functionals at ghost number one determines potential anomalies in the theory \cite{Howe:1990pz, Vandoren:1996ku, Howe:2006si}.

\section{The AKSZ construction of the A-model}
\label{sec:AKSZ_construction}

The space of fields and antifields in BV quantization naturally holds an odd Poisson structure, and the gauge fixing procedure requires the existence of a functional $S$ that obeys the master equation $(S, S)=0$.  We will call any supermanifold (finite or infinite dimensional) a P-manifold if it is equipped with an odd Poisson structure, and  a Q-manifold if it is equipped with an odd vector field 
\begin{equation}
\label{eq:Q_structure}
\widehat{Q} := ( \cdot, S)
\end{equation}
 that obeys $\widehat{Q}^2=0$ \cite{Schwarz:1992nx}. The latter is equivalent to the existence of a function(al) $S$ obeying $(S,S)=0$. The space of fields and antifields is an examples of an infinite dimensional PQ-manifold, that is, a supermanifold with both a P- and a Q-structure. An important example of a finite dimensional PQ-manifold is the tangent bundle of a Poisson manifold $\cM$ with parity reversed fibers, which we denote as $\Pi T \cM$. Taking  $X^i$ to be coordinates on $\cM$ and $\pi_i$ the odd coordinates on the fibers, the odd symplectic bracket is given by
\begin{equation}
(A, B) = \frac{\delr A}{\partial X^i} \frac{\dell B}{\partial \pi_i} -   \frac{\delr A}{\partial \pi_i} \frac{\dell B}{\partial X^i}  \ ,
\end{equation}
and the Q-structure is obtained from
\begin{equation}
S = \frac{1}{2} P^{ij}(X) \pi_i \pi_j \ .
\end{equation}
The master equation is satisfied if $P^{ij}$ defines a Poisson structure on $\cM$, i.e. if $P^{[ ij}_{ \ \ ,k} P^{m ]k} = 0$. 

In the AKSZ construction \cite{Alexandrov:1995kv} first a finite dimensional PQ-manifold $\cP$ is constructed, and then one makes use of a canonical PQ-structure on the space of maps from $\hatSigma$ to $\cP$, where $\hatSigma = \Pi T \Sigma$, and $\Sigma$ is an $n$-dimensional worldvolume. The Q-structure defines the extended action of the topological model. In the standard BV construction one starts from some classical physical action, so it is clear from the outset what the fields and the antifields are. This is no longer the case in the AKSZ construction, instead the choice forms a part of the definition of the theory. Geometrically it corresponds to a choice of Lagrangian submanifold of the PQ-manifold, that is, a submanifold with half the maximal dimension such that the odd symplectic form restricted to it vanishes. In the standard approach, in order to respect the classical limit, it is crucial that the physical fields have ghost number zero. This requirement can be dropped for topological theories, which in particular means that it is not a priori necessary to introduce auxiliary pairs, and that there is no need to restrict observables to have ghost number zero.

The target space for the A-model is the PQ-manifold $\Pi T \cM$, where $\cM$ is a K\"{a}hler manifold, and for the B-model the target space is $\Pi( T^{(0,1)} \cM \oplus T^{*(1,0)} \cM)$ \cite{Alexandrov:1995kv}. The quantum consistency of the $B$ model requires $\cM$ to be Calabi-Yau, while the target space of the A-model can be any K\"{a}hler manifold. The A- and B-models are special cases of a $\cJ$-model, which can be defined, at the classical level, on any generalized complex manifold  \cite{Pestun:2006rj}. For the A- and B-models the generalized complex structure in question reduces to, respectively,  $\Jsym$ and  $\Jcpx$ in (\ref{eq:sym_and_cpx}). The PQ-manifold is constructed directly from the Lie bi-algebroid structure (\ref{eq:Lie_algebroid}) that corresponds to the generalized complex structure in question. 

Let us take the local coordinates on $\Pi T \Sigma$ to be 
\begin{equation}
z := \{ \sigma^{+}, \sigma^-, \theta^+ , \theta^- \} \  ,
\end{equation}
where $\sigma^\pm$ are worldsheet light-cone coordinates and $\theta^\pm$ the fermionic fiber coordinates. The fields of the A-model, $X(z)$ and $\pi(z)$,  are referred to as \emph{de Rham superfields}. We note that $X(z)$ is a map  $\Pi T \Sigma \rightarrow \cM$, while $\pi(z)$ is a section of $ \Pi T^* \Sigma \otimes X^* ( \Pi T \cM)$. The P-structure is given by 
\begin{equation}
\label{eq:A_model_odd_symp_struct}
(A, B) :=   \int d^2 z \left( \frac{ \dr A} { \delta   X^i(z)  } \frac{\dl B}{ \delta \pi_i(z) } - 
\frac{\dr A}{ \delta \pi_i(z)}  \frac{ \dl B}{ \delta   X^i(z) }  \right)  \ ,
\end{equation}
and the AKSZ action for the A-model is
\begin{equation}
\label{eq:A_model_action}
S = \int d^2 z \left( \frac{1}{2}\omega^{ij}(X) \pi_i (z) \pi_j (z)+ \frac{1}{4} \omega_{ij} D X^i(z) DX^j(z) + \pi_i DX^i(z) \right) \ .
\end{equation}
where
\begin{equation}
d^2 z = d \sigma^+ d \sigma^-  d \theta^+ d \theta^- \ .
\end{equation}
$D$ is given by
\begin{equation}
D = \theta^+ \dep +  \theta^- \dem \ ,
\end{equation}
and obeys $D^2=0$, while $ \omega_{ij}$ is the  K\"{a}hler form and $ \omega^{ij}$ is its inverse.  From now on we will suppress the $z$ dependence of $X$ and $\pi$.

The action (\ref{eq:A_model_action}) has a ghost number symmetry, and we assign
\begin{equation}
\label{eq:A_model_gh_no}
\gh(X) = 0 \ \ \  , \ \ \  \gh (\pi) = -1 \  \ \  \mathrm{and} \ \ \  \gh(\theta^{\pm}) = 1 \ ,
\end{equation}
so that $\gh(D) = 1$, $\gh(d^2 z ) = -2$, and $\gh (S) = 0$. The de Rham superfields can be expanded in terms of the worldsheet superspace coordinates as
\begin{align}
\label{eq:superfield_exp}
& X^k = \phi^k + \theta^- \pi_*^{+k} + \theta^+ \pi_*^{-k} - \theta^+ \theta^- \chi^k_* \ , \\ \nonumber
& \pi_k = \chi_k + \theta^- \pi_{-k} + \theta^+ \pi_{+k} + \theta^+ \theta^- \phi^*_k \ ,
\end{align}
where we have made an initial choice which component fields are to be treated as fields and which as antifields.  The odd symplectic form is given by:
\begin{align}
\label{eq:poisson_antibracket}
(A,B) = &  \int d^2 z \left( \frac{ \dr A} { \delta   \phi^i } \frac{\dl B}{ \delta \phi^*_i} - \frac{\dr A}{ \delta \phi^*_i}\frac{ \dl B}{ \delta   \phi^i } 
 - \frac{ \dr A} { \delta   \chi_i } \frac{\dl B}{ \delta \chi^i_*} +  \frac{\dr A}{ \delta \chi^i_*} \frac{ \dl B} { \delta   \chi_i } \right.  \\ \nonumber
& \left. + \frac{ \dr A} { \delta   \pi_{- i} } \frac{\dl B}{ \delta \pi_*^{- i} } - \frac{\dr A}{ \delta \pi_*^{- i} } \frac{ \dl B} { \delta   \pi_{- i} } 
 - \frac{ \dr A} { \delta   \pi_{+ i} } \frac{\dl B}{ \delta \pi_*^{+ i} }  + \frac{\dr A}{ \delta \pi_*^{+ i} } \frac{ \dl B} { \delta   \pi_{+ i} }  \right) \ .
\end{align}
Integrating over the fermionic coordinates in (\ref{eq:A_model_action}) we obtain the extended action:
\begin{align}
\label{eq:poisson_action}
S =& \int d^2 \sigma \left( \vphantom{\frac{1}{2}} \frac{1}{2} \omega_{i j} \dep \phi^i \dem \phi^j +  \pip{i} \dem \phi^i - \pim{i} \dep \phi^i + \omega^{ij} \pip{i} \pim{j} \right. \\ \nonumber
& \left. +\pi_*^{+k} \left[ \dep \chi_k - \omega^{ij}_{ \ \ ,k} \chi_i \pip{j} \right] + \pi_*^{-k} \left[ - \dem \chi_k + \omega^{ij}_{ \ \  ,k} \chi_i \pim{j} \right]  \right.\\ \nonumber
& \left.  \vphantom{\frac{1}{2}}  + \phi^*_j \omega^{ji} \chi_i - \frac{1}{2} \chi_*^l \omega^{ij}_{ \ \  ,l} \chi_i \chi_j + \frac{1}{2} \pi_*^{+l} \pi_*^{-m}  \omega^{ij}_{ \ \ , lm}  \chi_i \chi_j \right) \ .
\end{align}
In terms of the standard BV construction described in Section \ref{sec:BV_procedure}, the  first line corresponds to the classical action, $\chi_k$ are the ghost fields, and the gauge symmetry transformations can be read off from the terms linear in $\pi_*$ and $\phi^*$. The term quadratic in the antifields is present due to the algebra of these symmetries closing only on-shell.  

The action of the form (\ref{eq:A_model_action}), but with the $\omega_{ij}$ term set to zero and $\omega^{ij}$ taken to be a general Poisson structure, is known as the Poisson $\sigma$-model. The choice of Lagrangian submanifold in (\ref{eq:superfield_exp}) is in fact the correct one for the quantization of topological models related to deformation quantization  \cite{Cattaneo:1999fm, Cattaneo:2001ys, Cattaneo:2001bp, Schomerus:1999ug}. To obtain the gauge fixed action one needs to introduce auxiliary pairs and construct a gauge fixing fermion much like for standard gauge theories. For the $p$-brane generalizations of the  Poisson $\sigma$-model see \cite{Park:2000au}.

Auxiliary pairs are not needed to define the A-model, but it is necessary to make a more refined choice of a Lagrangian submanifold than (\ref{eq:superfield_exp}) by making use of an almost complex structure on the target manifold (in what follows we take this structure to be actually complex),
\begin{align}
\label{eq:A_model_gauge_fixing}
\pi_{+ \alpha} & \rightarrow \psi^-_{* \alpha}   &  \pi_{+ \balpha} & \rightarrow   \pi_{+ \balpha}   
&  \pi_{ - \balpha} & \rightarrow \psi^+_{* \balpha}   & \pi_{- \alpha} &  \rightarrow \pi_{ - \alpha} \\ \nonumber
\pi^{+ \alpha}_*  & \rightarrow \psi_-^{\alpha}  &  \pi_*^{+ \balpha} & \rightarrow  \pi_*^{+ \balpha} 
& \pi_*^{- \balpha} & \rightarrow \psi_+^{\balpha}  &  \pi_*^{-\alpha} & \rightarrow \pi_*^{- \alpha} \ ,
\end{align}
referred to holomorphic coordinates. After making the replacements (\ref{eq:A_model_gauge_fixing}) in the extended action (\ref{eq:poisson_action}) and integrating out the auxiliary fields $\pi_{+\alpha}$ and $\pi_{- \bbeta}$ one obtains the extended action of the A-model\footnote{We are using the convention $\omega_{\alpha \bbeta} = - i g_{\alpha \bbeta}$, where $g_{ij}$ is the K\"{a}hler metric.}:
\begin{align}
\label{eq:A_model}
S  = & \int d^2 \sigma \left( \vphantom{\frac{1}{2}}  \frac{i}{2} g_{ij} \dep \phi^i \dem \phi^j + \chi_{\nu}(\dep \psi_-^\nu + \Gamma^\nu_{ \ \alpha \beta} \dep \phi^\alpha \psi_-^\beta) \right. \\ \nonumber 
& - \chi_{\bnu}(\dem \psi_+^\bnu + \Gamma^\bnu_{ \ \balpha \bbeta} \dem \phi^\balpha \psi_+^\bbeta)  - i R^{\alpha \bbeta}_{ \ \ \kappa \bnu} \psi_-^\kappa \psi_+^\bnu \chi_\alpha \chi_\bbeta  \\ \nonumber
& +i \phi^*_\nu g^{\nu \bkappa} \chi_\bkappa - i \phi^*_\bnu g^{\bnu \kappa} \chi_\kappa + \psi_{*\beta}^- ( \dem \phi^\beta  + i  \Gamma^{\beta \bbeta}_{ \ \ \alpha} \psi^\alpha_- \chi_{\bbeta}) \\ \nonumber
& - \psi_{*\bbeta}^+ ( \dep \phi^\bbeta  - i  \Gamma^{\bbeta \beta}_{ \ \ \balpha} \psi^\balpha_+ \chi_{\beta}) + i \chi_*^\alpha \Gamma^{\nu \bkappa}_{ \ \ \alpha} \chi_\nu \chi_\bkappa + i \chi_*^\balpha \Gamma^{\bkappa \nu}_{ \ \  \balpha} \chi_\nu \chi_\bkappa \\ \nonumber
& + i g^{\alpha \bbeta} \psi_{* \alpha}^- \psi_{* \bbeta}^+  \left. \vphantom{\frac{1}{2}} \right) \ .
\end{align}
In the AKSZ formulation the fields $\chi$ naturally have a downstairs index, whereas in the twisting procedure the index is naturally upstairs \cite{Alexandrov:1995kv, Bonechi:2007ar}. In the latter formulation one can associate $\chi^i$ with the differentials $dz^i$ on the target space, and one has simply $\delta \phi^i = \chi^i$ and $\delta \chi^i = 0$ under BRST transformations. In the AKSZ formulation $g^{ij} \chi_j$ is associated with $dz^i$, and this is the reason why $\chi_i$ doesn't transform trivially, as can be gathered from the presence of terms linear in $\chi_*$ in (\ref{eq:A_model}).

Finally, let us point out some subtleties in the AKSZ construction.  The A-model action obtained from the twisting procedure is characterized by the fact that its bosonic part is purely real, assuming that we take the action of the $(2,2)$ untwisted model to be real, due to the fact that the twisting leaves bosonic fields invariant. Our conventions differ by an inessential factor of $i$, so the bosonic part of  (\ref{eq:A_model}) comes out purely complex. Now, the second term in (\ref{eq:A_model_action}) is just the pullback of $\omega$ to the worldsheet
\begin{equation}
\label{eq:kahler_form_pullback}
 \int d^2 z \omega_{ij} D X^i DX^j = \int d^2 \sigma \omega_{ij} \dep \phi^i \dem \phi^j \ ,
\end{equation}
so it is topological, and the master equation is satisfied irrespective of its presence. However, its coefficient is determined if  we require the bosonic part of the action to be purely complex. There is a similar issue related to the role of a gauge fixing fermion. Following from (\ref{eq:A_model_gh_no}), the component fields have the ghost numbers
\begin{align}
\gh (\chi_i) = 1  \ \ \ ,  \ \  \gh (\psi_+^{\balpha}) =  \gh (\psi_-^\alpha) = -1 \  \ \ , \  \ \  \gh (\phi^i) = \gh (\pi_{+ \balpha }) = \gh ( \pi_{- \alpha}) = 0 \ ,
\end{align}
so a general gauge fixing fermion that is of second order in the fields, constructed using the K\"{a}hler metric, has the form:
\begin{equation}
\label{eq:A_model_gauge_fermion1}
\Psi = \int d^2 \sigma i g_{\alpha \bbeta} ( \dem \phi^\alpha \psi_+^{\bbeta} + \dep \phi^\bbeta \psi_-^\alpha)  \ .
\end{equation}
Starting from an action like (\ref{eq:A_model_action}), but with the $\pi_i DX^i$ term having a coefficient other than one, it would be necessary to perform a canonical transformation generated by a fermion proportional to (\ref{eq:A_model_gauge_fermion1}) in order to obtain the A-model action. However, if we start from precisely (\ref{eq:A_model_action}) no canonical transformation is necessary.

\subsection{Invariance of the A-model under $b$-transforms}
\label{sec:inv_under_b_transforms}

Under a $b$-transform (\ref{eq:b_transform}) the fields of the $A$-model transform as
\begin{equation}
\label{eq:AKSZ_b-transform}
X^i \rightarrow X^i \ \ \ \ \ \pi_i \rightarrow \pi_i +  b_{ij} DX^j  \ .
\end{equation}
One can check that this is a canonical transformation of the antibracket (\ref{eq:A_model_odd_symp_struct}) provided that $db=0$. The $A$-model action (\ref{eq:A_model_action}) is not invariant under (\ref{eq:AKSZ_b-transform}), but rather transforms to a special case of the Zucchini action  \cite{Zucchini:2004ta}:
\begin{equation}
\label{eq:Zucchini}
S  = \int d^2 z \left[  \left(- b_{ij} + \frac{1}{2} L_{ij} \right) DX^i DX^j + \pi_i DX^i + J^i_{ \ j} \pi_i DX^j + \frac{1}{2} P^{ij} \pi_i \pi_j \right] \ .
\end{equation}
The Zucchini action is a  solution to the master equation provided the tensors satisfy the first three conditions in Appendix \ref{app:integrability} with all the flux terms set to zero. Thus, the model can be defined on any manifold with a generalized complex structure, but the actual conditions coming from the master equation are weaker. In detail, we have
\begin{align}
\label{eq:zucchini_me}
(S, S) \equiv  \int d^2 z \left( S^{ijk} \pi_i \pi_j \pi_k+ V^{jm}_{ \ \ k} DX^k \pi_j \pi_m + T_{ij}^{ \ \ k} DX^i DX^j \pi_k \right) = 0\ .
\end{align}
The condition $W_{ijk} = 0$ doesn't feature because it involves a contraction with $(DX)^3$, which is identically zero.

For the $b$-transform of (\ref{eq:A_model_action}) we have
\begin{equation}
\label{eq:b_transformed_structure}
P^{ij} = \omega^{ij} \ \ \ , \ \ \  J^i_{ \ j} = \omega^{im} b_{mj} \ \ \ \mathrm{and} \ \ \  L_{ij} = \frac{1}{4} \omega_{ij} + \frac{1}{2} \omega^{mp} b_{im} b_{jp}  \ .
\end{equation}
Because (\ref{eq:AKSZ_b-transform}) is a canonical transformation, one can choose the same Lagrangian submanifold as for the A-model before the $b$-transform (see (\ref{eq:poisson_antibracket}) and (\ref{eq:A_model_gauge_fixing})).  Writing out  (\ref{eq:AKSZ_b-transform}) in components reveals that it  transforms both fields and antifields, so it is not a canonical transformation generated by a gauge fixing fermion that depends only on fields. However, it turns out that a canonical transformation generated by 
\begin{equation}
\label{eq:gf_fermion_b}
\Psi = \int d^2 \sigma \left(- b_{\alpha \bbeta} ( \dem \phi^\alpha \psi^{\bbeta}_+ + \dep \phi^\bbeta \psi_-^\alpha) 
 - b_{\balpha \bbeta} \dem \phi^\balpha \psi^\bbeta   + b_{\alpha \beta} \dep \phi^\alpha \psi_-^\beta \right)
\end{equation}
undoes the action of the $b$-transform, but only after integrating over the auxiliary fields $\pi_{+\alpha}$ and $\pi_{- \bbeta}$. If there was an obstruction to integrating out the auxiliaries, and we argue in Section \ref{sec:CY_with_flux} that this can happen when one introduces a gauge fixing fermion that depends on $H$-flux, then the impact of a transform by a closed $b$-field should be examined more carefully.

\section{The three dimensional A-type model}
\label{sec:3dAtype_model}

In this section we consider a class of topological membrane models that naturally include the $H$-, $f$-, $Q$- and $R$-fluxes that were introduced in (\ref{eq:courant_def}), and reduce to the Zucchini model (\ref{eq:Zucchini}) when the fluxes are turned off \cite{Ikeda:2007rn}. This is possible because the membrane is assumed to have a boundary, and in the absence of fluxes the AKSZ action becomes a total derivative. The Zucchini model itself reduces to the A-model when the generalized complex structure is given by $\Jsym$, but it does not reduce to the B-model when the generalized complex structure is $\Jcpx$ (see (\ref{eq:sym_and_cpx})). The reason why the A- and B-models can not be special cases of the same AKSZ action, at least for general dimension of the target space, is explained in  \cite{Pestun:2006rj}. B-type membrane models are also discussed in \cite{Ikeda:2007rn}. 

Let us first rewrite the standard A-model AKSZ action (\ref{eq:A_model_action}) in three dimensions, by acting on it with the $D$ operator. The maps from the worldsheet can be extended to maps from a membrane that has the worldsheet as its boundary, and the action obtained by integrating this Lagrangian over the membrane superfield coordinates is equivalent to the original A-model via Stokes' theorem.  To obtain a BV structure in three dimensions it is necessary to introduce the fields listed in Figure \ref{fig:AKSZ_membrane}.

\begin{figure}
\begin{center}
\begin{tabular}{ | c| c| c|}
\hline
 Field & Parity & Ghost number \\ \hline \hline
$A^i$	&	fermion & 1 \\ \hline
$B_i$ 	&     boson & 2 \\ \hline
$Y_i$	& boson & 2 \\	\hline
$Y_*^i$   & boson  & - 3 \\  \hline
$Z^i$	& fermion & 1 \\ \hline
$Z^*_i$ &   fermion & -2 \\ \hline
 \end{tabular}
 \end{center}
 \caption{The field content of three of the AKSZ membrane action.} \label{fig:AKSZ_membrane}
\end{figure}

The AKSZ membrane action is given by 
\begin{align}
\label{eq:3d_A_model}
S = &  \int d^3 y \left( \frac{1}{2} \omega^{ij}_{ \ \ ,k} A^k \pi_i \pi_j -  \omega^{ij} B_i \pi_j - \frac{1}{2} D X^i B_i + \frac{1}{2} D \pi_i A^i \right. \\ \nonumber 
 & \left. + (A^i - D X^i) Y_i + ( B_i + D \pi_i ) Z^i + O(Z^*) + O(Y_*)  \vphantom{\frac{1}{2}}  \right) \ ,
\end{align}
and the antibracket by
\begin{align}
\label{eq:3d_BV_bracket}
(F, G) =  &  \int d^3 y \left( \frac{ \dr F} { \delta   X^i } \frac{\dl G}{ \delta B_i}  -  \frac{ \dr F} { \delta   B_i } \frac{\dl G}{ \delta X^i}  +  \frac{ \dr F} { \delta   A^i } \frac{\dl G}{ \delta \pi_i} +  \frac{ \dr F} { \delta   \pi_i } \frac{\dl G}{ \delta A^i} \right. \\ \nonumber 
& \left. +  \frac{ \dr F} { \delta   Y_i } \frac{\dl G}{ \delta Y_*^i}  -  \frac{ \dr F} { \delta   Y_*^i } \frac{\dl G}{ \delta Y_i}  +  \frac{ \dr F} { \delta   Z^i } \frac{\dl G}{ \delta Z^*_i}  +  \frac{ \dr F} { \delta   Z^*_i } \frac{\dl G}{ \delta Z^i}  \right) \ .
\end{align}
As one can see, $B_i$ and $A^i$ play the role of antifields to $X^i$ and $\pi_i$, respectively. The field-antifield pairs have the same parity because the measure of the three dimensional superspace is fermionic. The superspace coordinates on the membrane are denoted by $y$, and $d^3 y$ is the fermionic measure. Eliminating the Lagrange multipliers $Y$ and $Z$ yields the total derivative action that reduces to the A-model on the boundary of the membrane.  In (\ref{eq:3d_A_model}) $O(Z^*)$ and $O(Y_*)$ denote terms that involve the antifields of $Z$ and $Y$, which we do not write down explicitly. Let us denote the part of the action in the first line of (\ref{eq:3d_A_model}) as $S_{(1)}$, and the part of the action in the second line, excluding the $O(Z^*)$ and  $O(Y_*)$ terms, as  $S_{(2)}$. Then  $(S_{(1)} , S_{(1)}) = 0$, provided that $\omega^{ij}$ is a Poisson structure, and  $(S_{(2)} , S_{(2)}) = 0$. However, $(S_{(1)} , S_{(2)}) \neq 0$, and one needs to introduce the $O(Z^*)$ and  $O(Y_*)$ terms in such a way the part of the full master equation coming from  $(S_{(1)} , S_{(2)})$ is cancelled. The completion of the extended action is straightforward but lengthy, and requires a repeated use of the Poisson structure condition. 

$S_{(1)}$ is a special case of an AKSZ membrane action that can be associated to a Courant algebroid on the bundle $T \cM \oplus T^* \cM$:
\begin{align}
\label{eq:Courant_top_model}
S_C = &  \int d^3 y \left( -\frac{1}{2} D X^i B_i + \frac{1}{2} D \pi_i A^i - f^{ij}_{(1)} B_i \pi_j - f^i_{(2) j} B_i A^j     \right.   \\ \nonumber 
 & \left. F_{(1) ijk} A^i A^j A^k + F_{(2) ij}^k A^i A^j \pi_k + F_{(3)k}^{ij} A^k \pi_i \pi_j  + F_{(4)}^{ijk} \pi_i \pi_j \pi_k  \vphantom{\frac{1}{2}} \right) \ .
\end{align}
The conditions for $S_C$ to satisfy the master equation can be grouped into three sets \cite{Roytenberg:2006qz, Ikeda:2006wd}:
\begin{align}
\label{eq:courant_1}
f^{(i}_{(2) m} f_{(1)}^{p)m} = 0  \ ,
\end{align}
\begin{align}
\label{eq:courant_2}
f_{(1) ,m}^{i [ j} f^{s ] m}_{(1)}  + F_{(3) m}^{js} f_{(1)}^{im} + 3 f_{(2)m}^{i} F_{(4)}^{jsm} & = 0  \\ \nonumber
f_{(1) , m}^{ij} f_{(2) s}^m - f_{(2) s ,m}^i f_{(1)}^{mj} + 2 f^i_{(2) m } F^{jm}_{(3) s} - 2 F_{(2) sm}^j f_{(1)}^{im} & = 0 \\ \nonumber
f_{(2)[ j, |m|}^{i} f_{(2) s] }^m - f_{(2) m}^i F_{(2) js}^m - 3 F_{(1) js m} f_{(1)}^{im} &  = 0  \ ,
\end{align}
and
\begin{align}
\label{eq:courant_3}
F_{(4) \  , m}^{[ijk} f_{(1)}^{ s ] m}  +3 F_{(3) m}^{[ij} F_{(4)}^{ks] m} & = 0 \\ \nonumber
F_{(3) k,m}^{[ij} f_{(1)}^{s]m}  + F_{(4) \ ,m}^{ijs} f_{(2) k}^m + 2 F_{(3) m}^{[ij} F_{(3)  \ k}^{s ] m}  - 6 F_{(2)km}^{[i} F_{(4)}^{js] m}  & = 0  \\ \nonumber
F_{(2) ijm}^{[k} f_{(1)}^{ s] m} - F_{(3) [ i , |m| }^{ks} f_{(2)j ] }^m + 9 F_{(1) ijm} F_{(4)}^{ksm} - 4 F_{(2)m[i}^{[k} F_{(3) \ j]}^{s]m} + F_{(3) m}^{ks} F_{(2) ij}^m & = 0  \\ \nonumber
- F_{(1) ijk, m} f_{(1)}^{ms} + F_{(2) [ ij , |m|}^s f_{(2)k]}^m + 6 F_{(1) [ij |m |} F_{(3) k]}^{sm}  +2 F_{(2)m [i}^s F^{m}_{(2) jk]} &= 0  \\ \nonumber
3 F_{(1) [ ij |m|} F^m_{ (2) pr]} - F_{(1) [ ijp, |m|} f_{(2) r]}^m & = 0  \ .
\end{align}
Including the $S_{(2)}$ term, one again needs to complete the extended action with $O(Z^*)$ and $O(Y_*)$ terms, and solving the full master equation involves a repeated use of the above conditions. The correspondence between AKSZ membrane actions and Courant algebroids (\ref{eq:def_Courant_alg})  was pointed out in \cite{Ikeda:2002wh, Ikeda:2002qx, Hofman:2002rv}, and has subsequently been studied extensively \cite{Roy1, Roytenberg:2006qz, Roytenberg:2002nu, Ikeda:2006wd}. The details of this correspondence for the bundle $T \cM \oplus T^* \cM$ are as follows:  $A^i$ and $\pi_i$ are coordinates along the fiber directions of $\Pi T \cM$ and $\Pi T^* \cM$ respectively,  the anchor map is given by 
\begin{align}
\pi (A^i) X^j =  f^{ij}_{(1)}  \ \ \ , \ \ \ \pi(B_i) X^j =  - f^j_{(2) i} \ , 
\end{align}
and there is an operation $\circ$ defined as
\begin{align}
\label{eq:courant_algebroid_sigma_model}
& A^i \circ A^j = - 2 F_{(3) k}^{i j} A^k - 6 F_{(4)}^{ijk} \pi_k \\ \nonumber
& A^i \circ \pi_j = - 2 F_{(2) j k}^i A^k + 6 F_{(3) j}^{i k} \pi_k \\ \nonumber 
& \pi_i \circ \pi_j =  -  6F_{(1) ijk} A^k -2 F_{(2) i j}^k \pi_k \ ,
\end{align}
from which the Courant algebroid bracket is obtained by antisymmetrization\footnote{One can check that the conditions (\ref{eq:def_Courant_alg}) are equivalent to (\ref{eq:courant_1}), (\ref{eq:courant_2}), and (\ref{eq:courant_3}). In fact, a Courant algebroid can be defined in two equivalent ways, either in terms of $[ , ]$ or $\circ$ \cite{Roy1}.}
\begin{equation}
\label{eq:courant_antisymmetrization}
[ e_1, e_2 ] = \frac{1}{2} ( e_1 \circ e_2 - e_2 \circ e_1 ) \ .
\end{equation}
The coefficients in the above equations were chosen to agree with the conventions in \cite{Ikeda:2006wd}. In an action corresponding to a Courant algebroid on a general product vector bundle $V \oplus V^*$  the $B$ fields would take values in the fiber coordinates of $T^* \cM$, while the $A$ and $\pi$ fields would take values in the fiber coordinates of $\Pi V$ and $\Pi V^*$ respectively. The action associated with a Courant algebroid  on a general vector bundle $E$ would contain the $B$ fields, and $A$ fields which take values in the fiber coordinates of $\Pi E$,  but it would not contain the $\pi$ fields \cite{Roytenberg:2002nu}. 

For the purpose of constructing a flux deformation of the Zucchini action the target space must be an almost generalized complex manifold. We therefore take
\begin{equation}
f^{ij}_{(1)}  = P^{ij}  \ \ \  \mathrm{and} \ \ \  f^i_{(2) j}  = J^i_{ \ j} \ ,
\end{equation}
and then equation  (\ref{eq:courant_1}) is just the second condition in (\ref{eq:gen_cpx2}). Furthermore, we wish to reproduce the integrability conditions of the almost generalized complex structure with respect to a Courant bracket deformed by not only $H$-flux, but also by $f$-, $Q$- and $R$-fluxes (\ref{eq:courant_def}). This is achieved by taking
\begin{align}
\label{eq:def_F_tensors}
F_{(1) ijk}  = &   \frac{1}{2} \left( L_{[ij,k] } +  \widetilde{H}_{ij k }   \right)   \ \ \ \ \ 
F_{(2) ij}^k  =    J^k_{ \ [ i , j] }   + \frac{1}{2} \widetilde{f}^k_{ \ ij}  \\ \nonumber
F_{(3)k}^{ij} = & \frac{1}{2} \left( P^{ij}_{ \ \ ,k}  + \widetilde{Q}^{ij}_{ \ \ k}  \right) \ \ \ \ \   \ 
F_{(4)}^{ijk} =  \frac{1}{2} \widetilde{R}^{ijk} \ ,
\end{align}
where
\begin{align}
\label{eq:def_F_tensors2}
 \widetilde{H}_{ij k }  = & - k_1 J^m_{ \ [i } H_{jk] m}  - k_2 L_{m [ i} f^m_{ \ jk ] } + k_2 L_{m [ i} J^p_{ \ j} J^s_{ \ k]} f^m_{ \ ps}  - k_3 L_{m [ k} L_{ | s | i} J^p_{ \ j]} Q^{sm}_{ \  \ p}   \\ \nonumber
 & - \frac{1}{3} k_4 L_{im} L_{jp} L_{ks} R^{mps} \\  \nonumber
  \widetilde{f}^k_{ \ ij}   = & k_1 P^{km} H_{m ij} + k_2 J^k_{ \ m} f^m_{ \ ij}  + 2 k_2 J^m_{ \ [i} f^k_{ \ j] m} - k_2 J^k_{ \ m} J^p_{ \ i} J^s_{ \ j} f^m_{ \ ps}  + 2 k_2 P^{m k} L_{ p [ i} J^s_{ \ j ]} f^p_{ s m} \\ \nonumber
  & -2 k_3 L_{p [ j} Q^{pk}_{ \ \ i]}  + 2 k_3 L_{m[i} J^p_{ \ j]} J^k_{ \ s} Q^{ms}_{ \ \ p} - k_3 P^{pk} L_{mi} L_{ s j} Q^{sm}_{ \ \ p}  - k_4 L_{im} L_{jp} J^k_{ \ s} R^{m p s} \\ \nonumber
\widetilde{Q}^{ij}_{ \ \ k}  = &    2 k_2 J^{ [ i}_{ \ m} P^{j] p} J^s_{ \ k} f^{m}_{ \ p s} - k_2L_{km} P^{pi} P^{sj} f^m_{ \ p s}  -2  k_2 P^{m [ j} f^{ i]}_{ \ mk} - k_3 J^m_{ \ k} Q^{ij}_{ \ \ m}  \\ \nonumber
& + k_3 J^{ [ i}_{ \ m} J^{ j]}_{ \ p} J^{s}_{ \ k} Q^{mp}_{ \ \ s}  - 2 k_3 L_{km} J^{[i}_{ \ p} P^{j ] s} Q^{m p}_{ \ \ s} - 2 k_3 J^{[ i}_{ \ m} Q^{j ]  m}_{ \ \ \  k} \\ \nonumber
& - k_4 J^i_{ \ m} J^j_{ \ p} L_{ sk} R^{mps} + k_4 L_{mk} R^{ijm} \\ \nonumber 
  \widetilde{R}^{kij}  = &  - k_2 J^{ [ k}_{ \ m} P^{i | p | } P^{j ] s} f^m_{ \ ps} + k_3 P^{m [ k} Q^{ i j ]}_{ \ \ \ m}  - k_3 J^{ [ k}_{ \ m} J^i_{ \ p} P^{j ] s} Q^{mp}_{ \ \ s} \\ \nonumber
& - \frac{k_4}{3} J^k_{ \ m} J^i_{ \ p} J^j_{ \ s} R^{mps} + k_4 J^{ [ k}_{ \ m} R^{ij] m}  \ .
\end{align}
  The conditions in (\ref{eq:courant_2})  become the first three conditions in Appendix \ref{app:integrability}. It is not obvious how the higher order conditions (\ref{eq:courant_3}) should be obtained from the deformed Courant bracket, except for the $H$-twisted case, when  satisfying (\ref{eq:courant_2}) requires $dH=0$, which is also the condition obtained by requiring the $H$-twisted bracket to be a Courant algebroid bracket.  For the particular examples we discuss in the next section (\ref{eq:courant_2}) express differential conditions that resemble Bianchi identities.  For standard and $H$-twisted generalized complex structures, the Courant algebroid structure defined by the topological membrane, (\ref{eq:courant_algebroid_sigma_model}), is related to the Lie bi-algebroid structure referred to after (\ref{eq:Lie_algebroid}), by a  correspondence between Lie bi-algebroid and Courant algebroid structures given in \cite{Gualtieri:2003dx}. This in particular means that the Courant bracket defined by the antisymmetrization of $\circ$ (\ref{eq:courant_antisymmetrization}) \emph{doesn't} correspond to the standard Courant bracket on $T \oplus T^*$ (\ref{eq:Courant_bracket}).

Unless the fluxes are turned off, $S_C$ is not a total derivative after the Lagrange multipliers $Y$ and $Z$ are integrated out. For the case with $H$-flux only it is still possible to work with a two dimensional model on an $H$-twisted generalized complex target space.  When the $b$-field is not closed, the odd symplectic structure (\ref{eq:A_model_odd_symp_struct}) gets deformed to 
\begin{equation}
\label{eq:deformed_BV_bracket}
(F, G)_H = (F, G) - \int d^2 z DX^i H_{ijk} \frac{\dl F}{\delta \pi_j} \frac{ \dl G}{\delta \pi_k}
\end{equation}
under a $b$-transform (\ref{eq:AKSZ_b-transform}), where we take $H = 3db$, and the master equation for the Zucchini action (\ref{eq:Zucchini}) with respect to $( , )_H$ reads
\begin{align}
 \int d^2 z \left( S^{ijk} \pi_i \pi_j \pi_k+ V^{jm}_{ \ \ k} DX^k \pi_j \pi_m + T_{ij}^{ \ \ k} DX^i DX^j \pi_k \right) = 0 \ ,
\end{align}
with $S$, $V$, and $T$ in Appendix  \ref{app:integrability} evaluated with $k_1 = 1$ and $k_2 = k_3 = k_4 = 0$. So, the master equation is satisfied with respect to the $H$-twisted bracket if the target space carries an $H$-twisted generalized complex structure, but as in the untwisted case (\ref{eq:zucchini_me}), the conditions coming from the master equation are weaker since $W_{ijk}=0$ doesn't feature. The problem when working with the twisted bracket is that one meets a rather serious technical difficulty in implementing the gauge fixing. Namely, it is completely unclear how to pick out a Lagrangian submanifold with respect to the odd symplectic structure (\ref{eq:deformed_BV_bracket}), due to the fact that  $(\pi_i, \pi_j)_H \neq  0 $,  so $X$ and $\pi$ are no longer conjugate variables.\footnote{Expanding $(S, S)_H$ in components reveals that the parts that come from the $H$-dependent deformation of the usual antibracket all depend on the auxiliary fields. This leaves some hope that one may be able to use the same Lagrangian submanifold as in the case without flux, and that the extended action obtained after integrating out the auxiliaries may satisfy the master equation with respect to the standard antibracket. One can easily check that this doesn't work. If we close our eyes and perform the gauge fixing as in the standard A-model, after performing a $b$-transform with $3db = H$ the extended action of the standard A-model $S_A$ (\ref{eq:A_model}) transforms to
\begin{align}
S = S_A +&  \int d^2 \sigma i \left( - H^{ \alpha}_{ \ \beta \nu} \psi_-^\nu \chi_\alpha \dep \phi^\beta 
+ H^\alpha_{ \ \beta \bnu} \psi_+^{\bnu} \chi_\alpha \dem \phi^\beta  
+  H^{ \balpha}_{ \ \bbeta \nu} \psi_-^\nu \chi_\balpha \dep \phi^\bbeta 
- H^\balpha_{ \ \bbeta \bnu} \psi_+^{\bnu} \chi_\balpha \dem \phi^\bbeta  \right. \\ \nonumber
& \left. - H^{ \alpha}_{ \ \bbeta \nu} \psi_-^\nu \chi_\alpha \dep \phi^\bbeta 
+ H^\alpha_{ \ \bbeta \bnu} \psi_+^{\bnu} \chi_\alpha \dem \phi^\bbeta  
+H^{ \balpha}_{ \ \beta \nu} \psi_-^\nu \chi_\balpha \dep \phi^\beta    
- H^\balpha_{ \ \beta \bnu} \psi_+^{\bnu} \chi_\balpha \dem \phi^\beta  \right) \ .
\end{align}
Thus, the change is proportional to $H$ and occurs only in the part of the action independent of antifields. It is easy to check that  this extra contribution isn't invariant under the BRST transformations of the A-model, and therefore the master equation with respect to the standard antibracket is not satisfied in the part independent of antifields.} The great advantage of the AKSZ membrane models is that this problem is avoided, since the standard BV bracket (\ref{eq:3d_BV_bracket}) is used.  In particular, for flux deformations of the A-model one can use the same Lagrangian submanifold as for the three dimensional rewriting of the standard A-model (\ref{eq:3d_A_model}).

We do not have a generalization of the twisted Zucchini model in two dimensions when the $f$-, $Q$- and $R$-fluxes are turned on, related to the fact that these fluxes can not be obtained via some generalization of the $b$-transform.  Two dimensional AKSZ actions have a correspondence with Lie algebroids (\ref{eq:Lie_algebroid})  \cite{Levin:2000fk, Olshanetsky:2002ur, Ikeda:2006wd}. As we mentioned,  the $\mathcal{J}$-models of \cite{Pestun:2006rj} are constructed from Lie bi-algebroids, and in \cite{Gualtieri:2003dx}  it is shown that one can always construct a Courant algebroid  from a Lie bi-algebroid. At the level of AKSZ this corresponds to the membrane action for the A-type model without flux becoming a total derivative. The fact that in the presence of $f$- $Q$- and $R$-fluxes the membrane actions can not be reduced to two dimensions indicates that the corresponding Courant algebroids can not be constructed from Lie bi-algebroids.

\section{Topological models on almost complex manifolds}
\label{sec:almost_cpx_models}

In this section we consider examples of $U(m)$ and $U(m) \times U(m)$  structure manifolds with flux which are good target spaces for the type of topological membrane model discussed in the previous section.  We will in particular be interested in geometries for which the Nijenhuis tensor of the almost complex structure(s) doesn't vanish.  There are two types of membrane models one can consider for each structure group. For $U(m)$ one has models associated with  $\Jsym$, and  $\Jcpx$, and  for $U(m) \times U(m)$ models associated with $\Jpm$ (see (\ref{eq:sym_and_cpx}) and (\ref{eq:bi_hermitian_gen_cpx0})).  We will mainly be interested in  $\Jsym$ and $\Jm$,  since these are the flux generalization of the A-model, while $\Jcpx$ and $\Jp$ are not related to the B-model \cite{Pestun:2006rj}.  In the previous section it was shown the AKSZ membrane action satisfies the master equation provided that the generalized complex structure in question is integrable with respect to the deformed Courant bracket (\ref{eq:courant_def}), and provided that the differential constraints on the fluxes (\ref{eq:courant_3}) are satisfied.  First we will examine geometries inspired by $(1,1)$ supersymmetric $\sigma$-models, for which the torsion of the connection that preserves the almost complex structure(s) is related to the $H$-flux, and the almost complex structures are responsible for additional supersymmetries of the $\sigma$-model.  When the Nijenhuis tensor of the almost complex structures doesn't vanish, for the $U(m)$ geometries one must  turn on the $f$-flux in the topological model, and for the $U(m) \times U(m)$ geometries the $H$-, $f$- and $R$-fluxes. The $f$- and $R$-fluxes are obtained by raising the indices of the $3$-form $H$.  We also consider $U(m)$ structure   manifolds for which the torsion is not related to $H$, and is not totally antisymmetric.   We show that all the conditions on the fluxes are satisfied for the $\sigma$-model inspired geometries, while for the more general $U(m)$ cases we obtain additional constraints on the torsionful Riemann tensor, and show that these are satisfied for three explicit examples of half-flat manifolds. 

The action for the $(1,1)$-supersymmetric $\sigma$-model is given by
\begin{equation}
S=\int\, d^2 z\, (g_{ij}+b_{ij}) D_+ X^i D_- X^j\ , 
\end{equation}
where $X$ is a map from $(1,1)$ superspace with coordinates $z = \{ \sigma^+, \sigma^-, \theta^+, \theta^- \}$ to a chart in the target space manifold, $g$ is the metric on the target space, $3db = H$, and $D_+$ and $D_-$ are supercovariant derivatives obeying:
\begin{equation}
D_+^2 = i\depp \  \ \ D_-^2 =  i \demm \ \ \ \{ D_+ , D_- \} = 0 \ .
\end{equation}
Second supersymmetries in the $(1,1)$ formulation take the form
\begin{equation}
\label{eq:second_susy}
\delta_{I_+} X^i = \varepsilon^{+}  I^{ \ \ i}_{(+) j}  D_+ X^j \ \ \ \ \ \delta_{I_-} X^i = \varepsilon^{-}  I^{ \ \ i}_{(-) j}  D_- X^j  \ ,
\end{equation}
with the conditions that $I^{ \ \ i}_{(+) j}$ and $I^{ \ \ i}_{(-) j} $ are almost complex structures and that they are covariantly constant with respect to two torsionful connections,
\begin{equation}
\nabla^{(\pm)} I_{(\mp)} = 0 \ ,
\end{equation}
whose  respective connection coefficients are
\begin{equation}
\label{eq:pm_connection}
\Gamma^{i ( \pm) }_{ \ jk} : =  \Gamma^{i }_{ \ jk} \pm \frac{1}{2} H^i_{ \ jk} \ .
\end{equation}
The conditions on the parameters $\varepsilon^{\pm}$ for the transformations (\ref{eq:second_susy}) to be symmetries are
\begin{equation}
D_{\mp} \varepsilon^{\pm} = 0 \ ,
\end{equation}
so $\varepsilon^{\pm}$ can be taken to depend on half the worldsheet superspace coordinates. 

If one requires the algebra of the second supersymmetries to be the standard $N=(2,2)$ algebra, $I_{(\pm)}$ are required to be complex and the model is said to have $(2,2)$ supersymmetry.\footnote{See \cite{Lindstrom:2004cd, Lindstrom:2004iw, Lindstrom:2004eh, Lindstrom:2006ee} for a formulation of $(2,2)$ $\sigma$-models with manifest $(1,1)$ supersymmetry that involves auxiliary superfields and is more natural in the context of the $T \cM \oplus T^* \cM$ bundle. A formulation with manifest $(2,2)$ supersymmetry in the context of generalized geometry is given in \cite{Lindstrom:2004hi, Lindstrom:2005zr, Bredthauer:2006hf, Lindstrom:2006ee, Lindstrom:2007qf, Lindstrom:2007xv, Lindstrom:2007sq}.}  The geometries that we want to consider are characterized by at least one almost complex structure, and this implies that the $N= (2,2)$ algebra is deformed by a non-linear symmetry related to the Nijenhuis tensor \cite{Delius:1989fy}. Schematically the $(+)$ sector of the standard $N=(2,2)$ algebra can be written as:
\begin{align}
\{ \delta_{I_+}, \delta_{I_+} \} \propto \delta_g  \ \ \ , \ \ \  \{ \delta_{I_+}, \delta_{g } \} \propto \delta_{I_+} \ \ \ , \ \ \   \{ \delta_{g }, \delta_{g } \} \propto \delta_g \ ,
\end{align}
where $\delta_g$ is the superconformal transformation,
\begin{equation}
\delta_g X^i = \varepsilon_g^{++} \depp X^i - \frac{i}{2} D_+  \varepsilon_g^{++}  D_+ X^i \ .
\end{equation}
The deformation enters in the $\{ \delta_{I_+}, \delta_{I_+} \}$ commutator
\begin{equation}
\{ \delta_{I_+}, \delta_{I_+} \} \propto \delta_g  + \delta_{N_+} \ ,
\end{equation}
where
\begin{equation}
\label{eq:nij_sym}
\delta_{N_+} X^i = \varepsilon_N^{++} N^i_{ (+) jk} D_+ X^j D_+ X^k \ ,
\end{equation}
and $N_{(+)}$ is the Nijenhuis tensor associated with $I_{(+)}$.   $N_{(+)}$  is covariantly constant with respect to $\nabla^{(-)}$, which follows from the closure of the algebra and can also be verified explicitly \cite{Delius:1989fy, Stojevic:2006pq}.  Replacing the partial derivatives in the Nijenhuis tensor using the covariant constancy of $I_{(+)}$  yields,  
\begin{equation}
\label{eq:Nijenhuis}
N_{(+) ijk} = H_{ijk} - 3 I^p_{  (+) [ i} I^s_{ (+)  j} H_{k ] p s} \ ,
\end{equation}
which just expresses that $N_{(+) ijk}$ is the $(3,0) + (0,3)$ component of $H$ with respect to $I_{(+)}$. Analogous equations can be written in the $(-)$ sector, and the symmetries in the $(+)$ and $(-)$ sectors commute. 

When $H$-flux is turned on it is possible to include only one second worldsheet supersymmetry, which means that the target space admits an almost complex structure covariantly constant with respect to, say, only $\nabla^{(+)}$, but no almost complex structure covariantly constant with respect to $\nabla^{(-)}$. Such models are called $(2,1)$ $\sigma$-models when the almost complex structure is actually complex. To distinguish the different possibilities, we introduce the following notation. For $(2^*, 1)$ models there exists a second supersymmetry corresponding to an almost complex structure $I_{(-)}$ such that $\nabla^{(+)} I_{(-)}=0$, and the algebra in the $(-)$ sector is deformed by the  Nijenhuis tensor symmetry (\ref{eq:nij_sym}). For $(2^*, 2)$ models there are two second supersymmetries, one corresponding to a non-integrable $I_{(-)}$, and the other to an integrable $I_{(+)}$. For $(2^*, 2^*)$ models neither $I_{(-)}$ nor $I_{(+)}$ are integrable.

An interesting point about the covariant constancy of $N$ is that it implies the reduction of the structure group. Consider, for example, the $(2, 1)$-model in three complex dimensions. The covariant constancy of $I$ implies that the structure group is reduced from $SO(6)$ to $U(3)$. For the $(2^*, 1)$-model  the covariant constancy of $N$ implies that the structure group is automatically reduced to $SU(3)$. Furthermore, for $n > 3$ the structure group reduces to a product group \cite{Stojevic:2006pq}. 

From the invariant tensors of a geometry with $U(m)$ structure one can construct two almost generalized complex structures $\Jcpx$ and $\Jsym$ (\ref{eq:sym_and_cpx}). The question we need to answer in order to construct the topological models is: what are the constraints on the parameters in (\ref{eq:courant_def}) such that $\Jcpx$ and $\Jsym$ are integrable with respect to $[ , ]_D$, provided that  the $U(m)$ tensors are covariantly constant with respect to some torsionful connection $\nabla^{(T)}$? It turns out that when $I^{i}_{ \ j} $ is almost complex  $\Jcpx$  and  $\Jsym$ are integrable with respect to (\ref{eq:courant_def}) with $k_2 = 1$, and all the other parameters set to zero, where we take the torsionful connection to be\footnote{ $\Gamma^{(sym)}$ is not equal to the Levi-Civita connection unless $T_{ijk}$ is totally antisymmetric. Rather $\Gamma^{(T)} = \Gamma^{L.C.} + \kappa$, 
where $\kappa$ is the contorsion, and the relation between contorsion and torsion is
\begin{equation}
\kappa^i_{ \ jk} = \frac{1}{2} ( T_{ \ jk}^i + T_{j \ k}^{ \ i} + T_{i \ k}^{ \ j} ) \  \ \ \ \ T^i_{ \ jk} = 2 \kappa^i_{ \ [ jk]} \ .
\end{equation}
}
\begin{equation}
\label{eq:torsionful_conn}
\Gamma^{i (T)}_{ \ jk} = \Gamma^{i (sym)}_{ \ jk} + \frac{1}{2} T^{i}_{ \ jk} \ ,
\end{equation}
and we associate
\begin{equation}
f^i_{ \ jk} = T^i_{ \ jk} \  
\end{equation}
in (\ref{eq:courant_def}). We are essentially twisting the original Courant bracket by torsion. When the torsion is totally antisymmetric, as is the case for $H$ in the $\sigma$-model described above, one needs to take $T^i_{ \ jk} =  -H^i_{ \ jk}$ when $I$ is covariantly constant with respect to $\nabla^{(-)}$. Of course, $\Jcpx$  and $\Jsym$  are also integrable with respect to this bracket when $I^{i}_{ \ j} $ is actually complex. For the complex case with torsion obtained from $H$, it turns out that one can consider a more general deformed bracket. Namely, $\Jcpx$  is actually integrable with respect to a Courant bracket with arbitrary parameters in (\ref{eq:courant_def}) turned on, and $\Jsym$ is integrable when $k_4 = 1$, $k_3 = -k_1$, and  $k_2 = 0$.  The $f$- $Q$- and $R$-fluxes are all obtained by raising the indices of the three-form $H$.

In bi-Hermitian geometry \cite{Gates:1984nk}, which corresponds to the $(2,2)$ $\sigma$-model,  a pair of complex structures are present, $I_{(\mp)}$, covariantly constant with respect to $\nabla^{(\pm)}$ (\ref{eq:pm_connection}).  As mentioned in Section \ref{sec:gen_cpx_geom},  this geometry can be expressed in terms of two generalized complex structures (\ref{eq:bi_hermitian_gen_cpx0}),
\begin{equation}
\label{eq:bi_hermitian_gen_cpx}
 \Jpm= \left ( \begin{array}{ll}
                      I^{ \ \ i}_{(+) j} \pm I^{ \ \ i}_{(-) j}  &  -(I_{(+)}^{ij} \mp I_{(-)}^{ij})\\
                      I_{ij}^{(+)} \mp I_{ij}^{(-)}    & -  ( I^{ (+) i}_{  \ \ j} \pm  I^{ (-) i}_{ \ \  j})
                    \end{array} \right ) \ ,
\end{equation}
which are integrable with respect to the standard $H$-twisted bracket (\ref{eq:H_twisted_courant}). If one considers integrability conditions with respect to a deformed bracket (\ref{eq:H_twisted_courant}) it is possible to accommodate  a more weakly constrained geometry, namely one with either $I_{(+)}$ or $I_{(-)}$ almost complex.  If we take $I_{(+)}$ to be almost complex, $ \Jpm$ are integrable with respect to the bracket (\ref{eq:courant_def}) with $k_1 = 1$, $k_2 = k_4= - \frac{1}{2}$, and when  $I_{(-)}$ is almost complex one needs to take  $k_1 = 1$, $k_2 = k_4= \frac{1}{2}$.  ${\cal J}_{(\pm)}$ are not integrable with respect to any deformed bracket  when both $I_{(+)}$ and $I_{(-)}$ are only almost complex. The fluxes are again obtained by raising the indices of $H$. In this paper we will not consider $U(m) \times U(m)$ geometries with torsion that is not totally antisymmetric.

Next we consider the differential conditions on the fluxes for the $U(m)$ A-type topological model, that is, the one associated with $\Jsym$ integrable with respect to  (\ref{eq:courant_def}) with $k_2=1$, $f^i_{ \ jk}=T^i_{ \ jk}$, and all the other fluxes set to zero. In this case $F_{(2)}$ and $F_{(4)}$ vanish, and the only non-trivial constraint is  the one in the second line of  (\ref{eq:courant_3}), which simplifies to
\begin{equation}
\label{eq:SUn_constraint}
F_{(3) k,m}^{[ij} f_{(1)}^{s]m}  + 2 F_{(3) m}^{[ij} F_{(3) \ k}^{s ] m} = 0   \ ,
\end{equation}
with $F_{(3) k,m}$ defined in (\ref{eq:def_F_tensors}) and  (\ref{eq:def_F_tensors2}) with $P^{ij} = I^{ij}$, $L_{ij} = I_{ij}$, and $J^{i}_{ \ j} = 0$.  (\ref{eq:SUn_constraint}) can be expressed in terms of the Riemann tensor of the torsionful connection
\begin{equation}
R^{i (T)}_{ \ jkl} = \Gamma^{i (T)}_{\ lj,k} - \Gamma^{i (T)}_{\ kj,l} + \Gamma^{m (T)}_{\ lj} \Gamma^{i (T)}_{\ km} - \Gamma^{m (T)}_{\ kj} \Gamma^{i (T)}_{\ lm} \ 
\end{equation}
 as
\begin{equation}
3 I^m_{ \ [i} I^p_{ \ j} R^{(T)}_{ s] [m p k]} + I^p_{ \ k} I^r_{ \ [i} I^{v}_{ \ j} I^m_{ \ s ]} R^{(T)}_{ p  r v  m} = 0 \ , 
\end{equation}
which in a holomorphic frame  reads:
\begin{equation}
\label{eq:SUn_constraint2}
R^{(T)}_{ \btau [ \alpha \beta \gamma ] } = 0 \ \ \ , \ \ \   R^{(T)}_{ \tau [ \balpha \bbeta \bgamma ] } = 0  \ .
\end{equation}
By definition, on a  manifold of $SU(3)$-structure there exists an invariant holomorphic form $\Omega$
\begin{equation}
R^{\tau (T)}_{ \ [ \alpha | \beta \gamma | } \Omega_{ \mu \nu ] \tau} = 0 \ .
\end{equation} 
which is equivalent to
\begin{equation}
R^{(T)}_{ \btau [ \alpha \beta \gamma ] } + 2 R^{(T)}_{ [ \alpha \beta } g_{ \gamma ] \btau} = 0 \ ,
\end{equation}
This means that for the $SU(3)$ case (\ref{eq:SUn_constraint2}) is equivalent to 
\begin{equation}
R^{(T)}_{ [ \alpha \beta ]} = 0 \ .
\end{equation}
For the $\sigma$-model case the torsion is given by $H^i_{ \ jk}$, and $dH= 0$, so one has the identity:
\begin{equation}
\label{eq:H_Bianchi}
R^{(T)}_{i [ jkl ] } = \nabla^{(T)}_{ i} H_{ jkl }  \ .
\end{equation}
Then (\ref{eq:SUn_constraint2}) implies that the $(3,0)$ and $(0,3)$ components of the $H$- field are covariantly constant with respect to the torsionful connection, which is true, as we pointed out in the context of equation (\ref{eq:nij_sym}). 

For the $U(m) \times U(m)$ geometries (\ref{eq:bi_hermitian_gen_cpx}) based on the $\sigma$-model the analysis is more elaborate than for the $U(m)$ cases, but the end result is similar. Namely, the covariant constancy of the Nijenhuis tensors associated with $I_{(+)}$ and $I_{(-)}$ is sufficient to satisfy the differential conditions on the fluxes.

For $U(3)$ geometries for which the torsion is not totally antisymmetric the above argument is no longer valid. In what follows we demonstrate that (\ref{eq:SUn_constraint2}) is satisfied for three explicit examples of half-flat manifolds, all of which are described in detail in \cite{Ali:2006gd}. We remind the reader that a half-flat manifold is an $SU(3)$-structure manifold \cite{Chiossi:2002tw} for which the following torsion classes vanish:
\begin{equation}
\cW_1^- = \cW_2^- = \cW_4 = \cW_5 = 0 \ ,
\end{equation}
where the minus denotes the imaginary part, which is equivalently expressed as
\begin{equation}
I \wedge dI = d \Omega^- = 0 \ .
\end{equation}
These manifolds are in general neither complex ($\cW_1 = \cW_2 =0$) nor K\"{a}hler  ($\cW_1 = \cW_2 = \cW_3 = \cW_4 = 0$).

The first case in question is the Iwasawa manifold \cite{Cardoso:2002hd, ketse}, for which only $\cW_3$ is non-zero (so it is in fact complex), and the metric is given by:
\begin{equation}
\label{eq:iwasawa_metric}
ds^2 = \sum_{i=1}^4 (dx^i)^2 + (dx^5 + x^1 dx^4 - x^3 dx^2)^2 + (dx^6 - x^1 dx^3 - x^4 dx^2)^2 \ .
\end{equation}
$I_{ij}$ has the following form for all three geometries,
\begin{equation}
I = e^1 \wedge e^2 + e^3 \wedge e^4 + e^5 \wedge e^6 \ ,
\end{equation}
where for the Iwasawa manifold:
\begin{align}
& e^j = dx^j \ , \ j = 1, ... , 4 \ , \\ \nonumber
& e^5 = dx^5 + x^1 dx^4 - x^3 dx^2 \ , \\ \nonumber
& e^6 = dx^6 - x^1 dx^3 - x^4 dx^2 \ .
\end{align}
The second case, referred to as the $\{ \cW_1^+, \cW_2^+ \}$ Iwasawa manifold,  is obtained by making the replacement
\begin{equation}
(dx^5 + x^1 dx^4 - x^3 dx^2) \rightarrow (dx^5 - x^1 dx^4 + x^3 dx^2)
\end{equation}
in both (\ref{eq:iwasawa_metric}) and $e^5$, and has $\cW_3=0$ but $\cW_1^+$ and  $\cW_2^+$ non-zero. For the third   case $\cW_1^+$,  $\cW_2^+$, and $\cW_3$ are all non-zero, the metric is given by
\begin{equation}
ds^2 = \sum_{i=1}^4 (dx^i)^2 + (dx^5 - x^1 dx^4 + x^3 dx^2)^2 + (dx^6 - x^4 dx^2)^2 \ ,
\end{equation}
and
\begin{align}
& e^j = dx^j \ , \ j = 1, ... , 4 \ , \\ \nonumber
& e^5 = dx^5 - x^1 dx^4 + x^3 dx^2 \ , \\ \nonumber
& e^6 = dx^6 - x^4 dx^2 \ .
\end{align}

With this information the contorsion can be calculated by making use of the identity
\begin{equation}
\kappa^i_{ \ jk} = -\frac{1}{2} I^i_{ \ m} \nabla^{L.C.}_{j} I^m_{ \ k} \ .
\end{equation}
We were able to check explicitly that (\ref{eq:SUn_constraint2}) holds for all three of the above geometries.\footnote{The calculation was done using the grtensor II Maple package.} On the $\{ \cW_1^+, \cW_2^+ \}$ Iwasawa manifold in fact the weaker condition
\begin{equation}
R^{(T)}_{i [jkl]} = 0 \ 
\end{equation}
is satisfied.

Finally, let us make a few comments about compactifications of string theory in relation to the geometries that we have discussed. The $\sigma$-model geometries with torsion that have $SU(3)$ or $SU(3) \times SU(3)$ structure are of special interest, since they are the ones that have a chance of playing a role in compactifications of type II string theory to Minkowski space that leave $N=1$ supersymmetry in four dimensions. Such compactifications have been extensively studied from the supergravity point of view.\footnote{See \cite{Grana:2005jc} for a review.} If we consider the classification of type II $N=1$ $SU(3)$ vacua \cite{Grana:2004bg, Grana:2005sn}, the only solutions that can be seen from the $\sigma$-model side are those that have $(2,1)$ worldsheet supersymmetry.\footnote{To include RR fields one would need to consider the GS or Berkovits formulation of the string theory $\sigma$-model.}  One can argue that $(2^*, 1)$ geometries don't appear in the above classification because they in fact break all spacetime supersymmetry. This is substantiated from the worldsheet perspective, because spacetime supersymmetry is described by the spectral flow of the $N=2$ algebra, which is apparently no longer present when the $N=2$ algebra is deformed by the Nijenhuis tensor symmetry (\ref{eq:nij_sym}). Thus one would expect $(2^*, 2)$ models to preserve $N=1$ spacetime supersymmetry and $(2^*, 2^*)$ models to again break all supersymmetry.\footnote{We note that Type II $SU(3) \times SU(3)$ vacua that preserve $N=1$ supersymmetry have not been classified in terms of torsion classes.} In addition to this, the Iwasawa manifold is a consistent background for compactifications of the Heterotic string to Minkowski space, since the requirement of spacetime supersymmetry is that the manifold is complex, $\cW_1 = \cW_2  = 0$, and that $2\cW_4 +\cW_5 = 0$  \cite{Cardoso:2002hd}.  The $\{ \cW_1^+, \cW_2^+ \}$ Iwasawa manifold is a consistent internal manifold for compactifications of type IIA string theory to $AdS_4$, since the condition on the intrinsic torsion for such compactifications is that it is contained in  $\cW_1^+ \oplus \cW_2^+$ \cite{Lust:2004ig}.

\section{Topological models on Calabi-Yau with three-form flux}
\label{sec:CY_with_flux}

The three dimensional AKSZ action (\ref{eq:Courant_top_model}) with only a closed $H$-flux turned on can always be reduced to two dimensions. This is due to the fact that one can always perform a $b$-transform with $db \propto H$, and choose the constant of proportionality in such a way that the $H$-twisted Courant bracket (\ref{eq:H_twisted_courant}) gets deformed to the standard Courant bracket, or equivalently, that the twisted BV bracket (\ref{eq:deformed_BV_bracket}) gets deformed to the untwisted one (\ref{eq:A_model_odd_symp_struct}). Let us consider the bi-Hermitian case as an example, when the two complex structures obey $\nabla^{\pm} I_{(\mp)} = 0$ (see (\ref{eq:bi_hermitian_gen_cpx})). One possibility to define an A-type topological model  is to start from the generalized structure ${\cal J}_{(-)}$, which is integrable with respect to the $H$-twisted bracket. To go down to two dimensions one can make use of the fact that 
\begin{equation}
{\cal J}_{(-)}' = \exp{(-b)} \Jm \exp{(b)}
\end{equation}
is integrable with respect to the standard Courant bracket (\ref{eq:Courant_bracket}). The three dimensional AKSZ action becomes a total derivative and reduces to the Zucchini action  (\ref{eq:Zucchini}) with
\begin{align}
& J^i_{ \ j} =  I^{ \ \ i}_{(+) j}  -  I^{ \ \ i}_{(-) j} + \left( I_{(+)}^{im} + I_{(-)}^{im} \right) b_{jm}   \ \ \  , \ \ \ P^{ij} = -\left( I_{(+)}^{ij} + I_{(-)}^{ij} \right)  \ , \\ \nonumber
& L_{ij} =  I_{ij}^{(+)} + I_{ij}^{(-)}  + \left( I^{ \ \ m}_{(+) j}  -  I^{ \ \ m}_{(-) j}  \right)b_{im} +  \left( I_{(+)}^{mp} + I_{(-)}^{mp} \right)b_{im} b_{jp} \ ,
\end{align}
where $3db = H$.  The $b$-field doesn't transform as a tensor unless $H$ is exact, and therefore neither do $J$ and $L$.  It would be interesting to gauge fix and compare this model with the topological model obtained by twisting the $(2,2)$ bi-Hermitian $\sigma$-model \cite{Kapustin:2004gv, Kapustin:2006ic, Zucchini:2006ii, Chuang:2006vt}.

Here we investigate a simpler case of a Calabi-Yau (CY) manifold with $H$-flux.   We emphasize that on a CY the complex structure $I$ and the $(3,0)+(0,3)$ form $\Omega$ are covariantly constant with respect to the Levi-Civita connection. By a CY with flux we mean that a three form $ H$ is turned on, but it doesn't play the role of torsion. The starting point  for the AKSZ construction of the A-model is the generalized structure $\Jsym$ (\ref{eq:sym_and_cpx}). In the presence of H-flux there are two possible types of deformations. One kind occurs at the level of the AKSZ action, and the other at the level of gauge fixing. The former is given by the membrane model defined by the generalized structure
\begin{equation}
\label{eq:b_transformed_symp1}
\Jsym' =  \exp{(b)} \Jsym \exp{(-b)} \ ,
\end{equation}
which is integrable with respect to the $H$-twisted bracket.  That is, taking
\begin{equation}
\label{eq:b_transformed_symp2}
 J^i_{ \ j}  = I^{im} b_{mj}  \ \ \ , \ \ \   P^{ij} = I^{ij} \ \ \ \mathrm{and} \ \ \  L_{ij} = I_{ij} + b_{im} b_{jp} I^{mp}
\end{equation}
in the action (\ref{eq:Courant_top_model}),  with $3db = H$, defines the natural three dimensional model on a CY with flux. 

A deformation can also occur at the level of gauge fixing, because one can pick a gauge fixing fermion that depends on a non trivial $b$-field, or on $H$ itself. In the remainder of this section we study such deformations for the standard A-model. In Section \ref{sec:BV_procedure} we showed that an anomaly-free theory should be invariant under deformations of the gauge fixing fermion,  but because the arguments leading to this conclusion are only valid at a perturbative level, they may fail for topological theories.\footnote{For an example of a construction that does depend on the gauge fixing fermion, in the context of one-dimensional topological theories introduced in \cite{Witten:1982im}, see  \cite{Rogers:2005zk}.} In the context of the standard A-model we showed that the role of the gauge fixing fermion (\ref{eq:gf_fermion_b}) was related to a $b$-transform when $db=0$. When $db \neq 0$ the model becomes the three dimensional one described above, and so the gauge fixing fermion  (\ref{eq:gf_fermion_b}) extended to a membrane plays a role in the definition of the three dimensional model based on (\ref{eq:b_transformed_symp1}).  It follows that the simplest deformation of the standard A-model is generated by the fermion constructed from the $(2,0) + (0,2)$ part of the $b$-field:
\begin{equation}
\label{eq:b_20_def}
\Psi_b = \int d^2 \sigma \left( b^{\alpha}_{ \ \bbeta} \pim{\alpha}  \psi_{+}^{\bbeta}  + b^{\bbeta}_{ \ \alpha} \pip{\bbeta} \psi_{-}^{\alpha}   \right) \ .
\end{equation}

The maps on which the A-model is evaluated are obtained by setting the BRST transformations of  $\psi_{+}^{\bbeta}$ and  $\psi_{-}^{\beta} $ to zero \cite{Witten:1991zz}. These can be read off from the terms proportional to  $\psi_{*\beta}^-$  and  $\psi_{*\bbeta}^+$ in (\ref{eq:A_model}), so:
\begin{align}
 \dem \phi^\beta  + i  \Gamma^{\beta \bbeta}_{ \ \ \alpha} \psi^\alpha_- \chi_{\bbeta} = 0  \ \ \ \ \ 
&  \dep \phi^\bbeta  - i  \Gamma^{\bbeta \beta}_{ \ \ \balpha} \psi^\balpha_+ \chi_{\beta} = 0 \ .
\end{align}
The solutions are obtained by setting the fermions to zero, and therefore the A-model is evaluated on holomorphic maps. After transforming the theory by the fermion (\ref{eq:b_20_def}) these maps are deformed to 
\begin{align}
\label{eq:holom_def}
\dem \phi^\alpha + i b^{\alpha \beta} \pim{\beta}  + i g^{\alpha \bbeta}_{ \ \ , \eta} \psi_-^\eta \chi_{\bbeta} - i b^{\balpha}_{ \ \tau} g^{\alpha \bbeta}_{ \ \ , \balpha} \psi_-^\tau \chi_{\bbeta}=0  \ ,
\end{align}
before the auxiliary fields $\pim{\beta}$ are integrated out (see Appendix \ref{app:A_model_w_aux}). There is a similar expression involving $\dep \phi^\bbeta$. The term in the action quadratic in the auxiliary fields is deformed to
\begin{equation}
\label{eq:pi_term_def}
-i \int d^2 \sigma \left( g^{\alpha \bbeta} + b^{\alpha}_{ \ \bkappa} b^{\bbeta}_{ \ \eta}  g^{\bkappa \eta} \right) \pip{\bbeta} \pim{\alpha} \ .
\end{equation}
The object in the parentheses needs to be inverted in order to obtain the equations of motion for the auxiliaries, and we can expect an inverse to  exist at a generic point in the field space. Let us denote the object in the parentheses as $\tilde{g}^{\alpha \bbeta}$ and its inverse as $\tilde{g}_{\alpha \bbeta}$. Then after eliminating the auxiliaries, we obtain that the model is evaluated on maps satisfying
\begin{align}
\label{eq:holom_maps1}
\dem \phi^\alpha - b^{\eta \alpha} \tilde{g}_{\eta \bnu} \dem \phi^\bnu + b^{\eta \alpha} \tilde{g}_{\eta \bnu} b^{\bnu}_{ \ \kappa} \dem \phi^\kappa = 0 \ .
\end{align}
When $b$ transforms like a tensor this is a contribution that one would get from an infinitesimal change of coordinates of the form
\begin{equation}
\phi^\alpha \rightarrow \phi^\alpha + A^\alpha_{ \ \beta} \phi^\beta + B^\alpha_{ \ \bbeta} \phi^\bbeta 
\end{equation}
which, due to the presence of the last term, doesn't respect the choice of complex structure. Since the A-model does not depend on the choice of complex structure,  transformations of this form will leave the theory invariant. However, when  $3db = H$  with $H$ not exact, the model is evaluated on maps which are not holomorphic with respect to \emph{any} complex structure on the target space.

How could this change the content of the theory? The partition function of the A-model coupled to topological gravity\footnote{If one doesn't couple the A-model to topological gravity only the genus zero contribution survives. For a review of topological strings the reader is referred to one of the following \cite{Neitzke:2004ni, Vonk:2005yv, Marino:2004uf}.} can be evaluated exactly at each genus $g$ and takes the form,
\begin{equation}
Z_g = \sum_\beta N_{g, \beta} \exp{(-t \omega \cdot \beta)} \ .
\end{equation}
Here $t$ is a measure of the size of the CY (which we have implicitly set to one in (\ref{eq:A_model})), $\omega$ is the K\"{a}hler form, and 
\begin{equation}
\omega \cdot \beta := \int_{\phi(\Sigma)} \omega \ ,
\end{equation}
where $\phi(\Sigma)$ denotes the homology class of the embedding of the worldsheet in the target space. It is straightforward to show that $\omega \cdot \beta $ is independent of the element of the homology class.  The exponential term is present because, up to terms proportional to equations of motion, the A-model action can be written in the form 
\begin{equation}
\label{eq:A_model_split}
S = t \omega \cdot \beta - it \int d^2 \sigma \{ Q, V \} \ ,
\end{equation}
where $Q$ is the A-model BRST operator.\footnote{The reason why the equivalence is only up to the equations of motion is that we are using the BRST operator instead of  the full BV operator (\ref{eq:BVoperator}).} Therefore, the partition function has two contributions, one from $\omega \cdot \beta $, that depends on the K\"{a}hler form, and the other which is a $Q$ of something. Naively the latter shouldn't contribute to the path integral (see the discussion around (\ref{eq:observables})). However, due to an anomaly there actually is a contribution, and it is given by the Gromov-Witten invariants $N_{g, \beta}$. These are topological invariants of the CY manifold that, in a certain sense, count the number of holomorphic maps in the target space at each genus and in each homology class. Going back to the deformation (\ref{eq:holom_maps1}) with a $b$-field that doesn't transform tensorially, one expects the numbers   $N_{g, \beta}$ to change, because on one hand a change in the gauge fixing fermion corresponds to a change of the second term in (\ref{eq:A_model_split}), and on the other hand the partition function is no longer evaluated on holomorphic maps. The $\omega \cdot \beta $ contribution clearly remains the same.

Finally, we give some remarks about the impact of gauge fixing fermions that depend on $H$-flux. The  possibilities are a lot more numerous compared with fermions that can be constructed from the $b$-field and the K\"{a}hler form, and can be classified according to the number of auxiliary fields they contain. Let us illustrate with an example from each class:
\begin{equation}
\int d^2 \sigma H_{i j \bbeta} \dem \phi^i \dem \dep \phi^j \psi_+^{\bbeta}   \ \ \ , \ \ \  \int d^2 \sigma H^\alpha_{ \ j \bbeta} \pim{\alpha} \dep \dem \phi^j \psi_+^{\bbeta}  \ \ \ , \ \ \  \int d^2 \sigma H^{\alpha \gamma }_{ \ \ \bbeta} \pim{\alpha} \dep \pim{\gamma} \psi_+^{\bbeta} \ .
\end{equation}
The case without auxiliaries deforms the holomorphic maps by $H$ dependent terms. For a single auxiliary the same is true, but, as for the fermion (\ref{eq:b_20_def}), one needs to invert the tensor that appears in the term quadratic in the auxiliaries, which will now depend on $H$. For the case with two auxiliaries it is a lot less clear how to proceed, since after the canonical transformation the action contains a term quartic in  the $\pi$ fields, in general with worldsheet derivatives acting on them. More work needs to be done to understand how to handle these cases. It may be that the $\pi$ fields need to be treated as propagating fields, in which case one would need to introduce auxiliary pairs and perform gauge fixing as for the Poisson $\sigma$-model (\ref{eq:poisson_action}).

\section{Conclusions}
\label{sec:conclusions}

In this paper we have studied a topological A-type models with flux in the AKSZ construction, arguing that generically such models are topological membranes rather than topological strings, and have shown that they can be defined on a large class $U(m)$- and $U(m) \times U(m)$-structure geometries. From the point of view of type II string theory compactifications both rank three fluxes and almost complex geometries are of interest in the context of mirror symmetry. This stems from the fact NS-NS $H$-flux plays a more intricate role in mirror symmetry than R-R fluxes.  Whereas R-R fluxes are simply interchanged, turning on $H$-flux has a more dramatic impact.  In \cite{Gurrieri:2002wz} it was shown that the mirror manifold of a Calabi-Yau three-fold with electric $H$-flux is a half-flat manifold. The flux in this case is not a three-form but is related to the intrinsic torsion. In \cite{Grana:2006hr} the analysis was extended to magnetic $H$-flux, and the mirrors were shown to be manifolds with $SU(3) \times SU(3)$ structure.  While we have shown that the topological membrane model can be defined on three particular examples of half-flat manifolds, it would be interesting to find out whether this is generically true, or whether there is a restriction on the possible half-flat manifolds.  It is worth stressing that the above examples are not solutions of string theory compactified to Minkowski space, so it is of some interest that the topological membrane can be defined on a class of almost complex geometries inspired by $(1,1)$ $\sigma$-models, which are expected to contain solutions that preserve $N=1$ supersymmetry. The half-flat geometries can be lifted to solutions of M-theory, and in this context it would be interesting to investigate whether  there is a relation between our construction and the membrane models of topological M-theory \cite{Dijkgraaf:2004te} studied in \cite{Bonelli:2005ti, Bonelli:2005rw, Bonelli:2006ph}.

Furthermore, mirrors of manifolds with $H$-flux are in general not expected to be geometric. Non-geometry has mostly been studied in the context of T-duality, for example, on a six-torus with $H$-flux supported on some three-cycle  \cite{Shelton:2005cf}. After performing a single T-duality along a direction with non-zero flux, one obtains a twisted torus, which is characterized by the tensor $f^i_{ \ jk}$. After preforming a second T-duality one obtains a manifold which is geometric only locally. It is still possible to understand it geometrically,  by doubling the directions supported by the $H$-flux, and considering transition functions between patches that include T-duality transformations; this type of geometry is referred to as a T-fold \cite{Hull:2004in, Hull:2005hk, Dabholkar:2005ve, Hull:2006va, Hull:2006qs}. A different way to understand this $T$-dual  is in terms of a manifold fibered by non-commutative tori \cite{Mathai:2004qc, Mathai:2004qq, Mathai:2005fd}. The various points of view have been reconciled in \cite{Grange:2006es}, and it is understood that non-commutativity is seen only in the open string sector \cite{Kapustin:2003sg, Grange:2007bp}. Performing a T-duality along all three directions supported by $H$-flux is conjectured to lead to a space which is not geometric even locally \cite{Bouwknegt:2004ap, Shelton:2005cf} (interestingly, \cite{Hull:2007jy} conjectures that it can still be understood in the T-fold formalism). Analogously to $H$ and $f$, the fluxes $Q^{ij}_{ \ \ k}$ and $R^{ijk}$ can be associated to the two more exotic spaces. The fluxes that arise when considering mirror symmetry of Calabi-Yau manifolds  with flux can be classified in the same way  \cite{Grana:2006hr}.  In light of all this, the fact that the fluxes in the membrane model  follow the same pattern is very suggestive, and in future work we would hope to obtain a more concrete understanding of the relations involved.

The analysis of T-duality transformations on tori with NS-NS $H$-flux has also revealed that  $\beta$-transforms, where $\beta$ is a $(0,-2)$ type tensor, are "mirror" to  a subset of $b$-transforms, where $b$ is a  $(0,2)$-form \cite{Kapustin:2003sg, Grange:2006es, Grange:2007bp}. The analysis suggests that non-commutativity on an $SU(3) \times SU(3)$-structure manifold with $Q$-type flux can be understood in terms of a deformation of the B-model action by a Poisson $\sigma$-model action, with the Poisson structure given by $\beta$. On the mirror side this is related to a deformation of the A-model on a Calabi-Yau with magnetic flux by  a non-trivial $(0,2)$ $b$-field, but it is not clear precisely how the deformation should arise, which suggests that perhaps the deformations we have discussed in Section \ref{sec:CY_with_flux} should be taken seriously in this context. One should note, however, that the mirror map for non-commutative deformations constitutes a very difficult problem. While the category of B-branes, including their non-commutative deformations, is well understood, the relation to the "mirror" category of A-branes is far from clear \cite{Kapustin:2001ij, Bressler:2002eu, Kapustin:2003sg, Kapustin:2005vs}. It has been suggested that  to clarify the relation one  should understand deformation quantization of  the Poisson structure associated with the A-model. As explained in Section \ref{sec:AKSZ_construction}, this requires keeping the auxiliary fields, which one is also forced to do when considering deformations of the A-model by gauge fixing fermions that depend on $H$, as we argued in Section \ref{sec:CY_with_flux}.

Finally, we expect that it would be rewarding to study topological models on bi-Hermitian geometries from the AKSZ perspective. These have been constructed via the twisting procedure \cite{Kapustin:2004gv, Kapustin:2006ic, Zucchini:2006ii, Chuang:2006vt}, but not all aspects are fully understood to date.

\section{Acknowledgements}

I would like to thank Jan Louis and Ron Reid-Edwards for many useful discussions. I would also like to acknowledge an ongoing collaboration with Paul Howe, which, while not on the topic of this paper, has inspired some of the results.  I am very grateful to Noriaki Ikeda, Andrei Micu, and Dmitry Roytenberg for useful and patient correspondence, and would like to acknowledge the helpful advice from Vasily Pestun and Thomas Strobl during the "Poisson Sigma Models, Lie Algebroids, Deformations, and Higher Analogues" workshop at the Erwin Schr\"{o}dinger International Institute for Mathematical Physics in Vienna. I am very grateful for the generous funding from the German Research Foundation (DFG).

\appendix

\section{Properties of the antibracket}
\label{app:poisson_antibracket}

\begin{align}
\epsilon ( A, B ) & =  \epsilon_A + \epsilon_B \ , \\
( A, B ) & =  - (-1)^{ ( \epsilon_A +1)( \epsilon_B +1) } ( B, A )  \ , \\
( A + B, C ) & =  ( A, C ) + ( B, C )  \ ,\\
(-1)^{(\epsilon_A+1)( \epsilon_C+1)} ( A, ( B, C ) ) + \mathrm{CYCLIC} &= 0 \ , \\  \label{eq:ab_jacobi}
( AB, C ) & =  A ( B, C ) + (-1)^{\epsilon_B (\epsilon_C+1)} ( A,C ) B \ .
\end{align}
Here $\epsilon_A \in \mathbb{Z}_2 $ is zero when $A$ is  a bosonic object and one when $A$ is fermionic.

\section{Integrability conditions for the deformed Courant Bracket}
\label{app:integrability}

Written below are the component expressions for the integrability conditions (\ref{eq:int_conditions}) of a an almost generalized complex structure (\ref{eq:gen_cpx_structure}) with respect to the deformed Courant bracket (\ref{eq:courant_def}).

\begin{align}
\label{eq:int_condition1}
S^{ijk} :=& P^{[ ij}_{ \  \ \ , m} P^{|m| k]} + k_2  P^{m[ k} P^{|p| j} f^{i]}_{ \ mp} + 2 k_3 P^{m[k} J^{j }_{ \ p} Q^{i] p}_{ \ \ m}  \\ \nonumber
& -\frac{k_4}{3} \left(  R^{ikj} - 3 J^{[k}_{ \ \ p} J^j_{ \ m} R^{i ] pm} \right) = 0 
\end{align}

\begin{align}
\label{eq:int_condition2}
V^{jm}_{ \ \ k} := & P^{i [ m} J^{j]}_{ \ i,k} - 2 P^{i [ m} J^{j ] }_{ \ k, i} + J^{i}_{ \ k} P^{ j m}_{ \ \ ,i} 
-P^{i [ m}_{ \ \ ,k} J^{j ]}_{ \ i} - k_1 P^{ij} P^{pm} H_{ipk}  \\ \nonumber
& + k_2 \left( -2  J^p_{ \ k} P^{i [ j} f^{m ]}_{ \ pi} - 2 J^{[ m}_{ \ \ p}  P^{j] i} f^p_{ \ ki} \right) \\ \nonumber
& + k_3 \left( -Q^{mj}_{ \ \ k} -2 J^p_{ \ k} J^{[j}_{ \ \ i} Q^{m] i}_{ \ \ \ p} + J^{m}_{ \ i} J^{j}_{ \ p} Q^{ip}_{ \ \ k} + 2 L_{ik} P^{p[m} Q^{j]i}_{ \ \ \ p} \right) \\ \nonumber
& +2 k_4 L_{ik} J^{[j}_{ \ \ p} R^{m]ip} = 0
\end{align}

\begin{align}
\label{eq:int_condition3}
 T_{ij}^{ \ \ k} := & 3 L_{[ij, m]} P^{mk} + 2 \left( J^k_{ \ m} J^{m}_{ \ [j, i]} + J^m_{ \ [j} J^k_{ \ i ],m} \right)
+2k_1 P^{mk}  J^{p}_{ \ [j} H_{i] mp} \\ \nonumber
 & + k_2 \left( f^k_{ \ ij} +2 P^{mk} L_{p [ j} f^p_{ \ i]m}   + 2 J^k_{ \ m} J^p_{ \ [ j}  f^m_{ \ i]p} - J^m_{ \ i} J^p_{ \ j} f^k_{ \ mp} \right) \\ \nonumber
 & + k_3 \left( 2  L_{m [i } J^p_{ \ j]} Q^{mk}_{ \ \ p}  + 2 J^k_{ \ p} L_{m [ j} Q^{mp}_{ \ \ i]}  \right) + k_4 L_{im} L_{pj} R^{mpk} = 0
\end{align}

\begin{align}
\label{eq:int_condition4}
 W_{ijk} := & 2 L_{kp} J^p_{ \ [ j, i ] }  + L_{p [ i} J^{p}_{ \ j],k} - 3 J^{p}_{ \ k} L_{[pj, i]}
+ J^p_{ \ [j} L_{ i] p, k} + 2 J^p_{ \ [i} L_{j ] k, p} \\ \nonumber 
& + k_1 \left( H_{ijk} - 3 J^p_{ \ [i} J^m_{ \ j} H_{k]pm} \right) \\ \nonumber
& +6  k_2 J^m_{ \ [i} L_{ j |p| } f^p_{ \ k ] m}  + 3 k_3 L_{m [k } L_{j | p | } Q^{pm}_{ \ \ i]} = 0
\end{align}

\section{A-model extended action with auxiliary fields}
\label{app:A_model_w_aux}

Below we give the action for the A-model with the auxiliary fields $\pi_+$ and $\pi_-$. It is the starting point for understanding deformations of the A-model of the kind described in section \ref{sec:CY_with_flux}.

\begin{align}
S  = & \int d^2 \sigma i \left[ \vphantom{\frac{1}{2}} - g_{\alpha \bbeta} \dep \phi^{\alpha} \dem \phi^{\bbeta}  + g_{\alpha \bbeta} \dep \phi^{\bbeta} \dem \phi^{\alpha} - g^{\alpha \bbeta} \pip{\bbeta} \pim{\alpha} \right. \\ \nonumber
& + \chi_\alpha \dep \psi^\alpha_-  - \chi_\balpha \dem \psi_+^\balpha  - \pim{\alpha}  \dep \phi^\alpha + \pip{\balpha} \dem \phi^\balpha   \\ \nonumber 
& -  g^{\kappa \bbeta}_{ \ \ , \balpha} \psi_{+}^{\balpha}  \chi_{\bbeta} \pim{\kappa}   +g^{\beta \bkappa}_{ \ \ , \alpha} \chi_{\beta} \pip{\bkappa} \psi_{-}^{\alpha}    + g^{\alpha \bbeta}_{ \ \ , \kappa \bnu} \psi_{-}^{\kappa} \psi_{+}^{\bnu} \chi_{\alpha} \chi_{\bbeta}   \\ \nonumber
& + \phi^*_{\alpha} g^{\alpha \bbeta} \chi_{\bbeta} -  \phi^*_{ \balpha} g^{\balpha \beta} \chi_{\beta} - \chi_*^{\alpha} g^{\beta \bkappa}_{ \ \ , \alpha} \chi_{\beta} \chi_{\bkappa} - \chi_*^{\balpha} g^{\beta \bkappa}_{ \ \ , \balpha} \chi_{\beta} \chi_{\bkappa} \\ \nonumber
&+  \psi_{* \bkappa}^{+} \left( i \dep \phi^\bkappa  + g^{\beta \bkappa}_{ \ \ , \balpha}  \psi_+^{\balpha} \chi_{\beta} \right) +  \psi_{* \kappa}^- \left( -i \dem \phi^\kappa +  g^{\kappa \bbeta}_{ \ \ , \alpha} \psi_-^{\alpha} \chi_{\bbeta} \right) \\ \nonumber
& -  \pi_*^{- \nu} \left( \dem \chi_\alpha +  g^{\bbeta \kappa}_{ \ \ , \nu} \chi_{\bbeta} \pim{\kappa} + g^{\alpha \bbeta}_{ \ \ , \kappa \nu} \psi_-^{\kappa} \chi_{\alpha} \chi_{\bbeta} \right)  \\ \nonumber 
&- \pi_*^{+ \balpha} \left(  - \dep \chi_{\balpha} + g^{\beta \bkappa}_{ \ \ , \balpha} \chi_{\beta} \pip{\bkappa} + g^{\alpha \bbeta}_{ \ \ , \balpha \bnu} \psi^{\bnu}_+ \chi_{\bbeta} \chi_{\alpha} \right) \\ \nonumber
& \left. +  g^{\alpha \bbeta} \psi_{* \alpha}^- \psi^+_{ * \bbeta} -  \pi_*^{- \alpha} \psi_{* \bkappa}^+ g^{\beta \bkappa}_{ \ \ , \alpha} \chi_{\beta}  - \pi_{*}^{+ \balpha} \psi_{* \kappa}^- g^{\kappa \bbeta}_{ \ \ , \balpha} \chi_{\bbeta} +  \pi_*^{+\bkappa} \pi_*^{ - \nu}  g^{\alpha \bbeta}_{ \ \ , \bkappa \nu} \chi_{\alpha} \chi_{\bbeta}   \vphantom{\frac{1}{2}} \right]
\end{align}


\end{document}